\documentclass[pdflatex,sn-mathphys]{sn-jnl}

\usepackage[nameinlink]{cleveref}
\usepackage{ragged2e}
\usepackage{astjnlabbrev}
\usepackage{color}
\usepackage{xspace}
\usepackage[version=4]{mhchem} 

\newcommand\eureka{\texttt{Eureka!}\xspace}
\newcommand\firefly{\texttt{FIREFLy}\xspace}
\newcommand{\smarter}{\texttt{smarter}\xspace}
\newcommand{\POSEIDON}{\texttt{POSEIDON}\xspace}
\newcommand{\planetname}{LHS~475b\xspace}

\DeclareSymbolFont{UPM}{U}{eur}{m}{n}
\DeclareMathSymbol{\umu}{0}{UPM}{"16}
\newcommand\micro{$\umu$}
\newcommand\micron{\micro m\xspace}
\newcommand\microns \micron

\usepackage[version=4]{mhchem} 

\jyear{2023}%

\begin{document}

\title[Lustig-Yaeger \& Fu et al. --- LHS~475b with JWST]
{A JWST transmission spectrum of a nearby Earth-sized exoplanet} 

%


\author*[1]{Jacob Lustig-Yaeger}\email{jacob.lustig-yaeger@jhuapl.edu}
\affil[1]{Johns Hopkins APL, Laurel, MD, 20723, USA}

\author*[2]{Guangwei Fu}\email{guangweifu@gmail.com}
\affil[2]{Department of Physics \& Astronomy, Johns Hopkins University, Baltimore, MD, USA}

\author[1]{E. M. May}

\author[3]{Kevin N. Ortiz Ceballos}
\affil[3]{Center for Astrophysics ${\rm \mid}$ Harvard {\rm \&} Smithsonian, 60 Garden St, Cambridge, MA 02138, USA}

\author[4]{Sarah E. Moran}
\affil[4]{Lunar and Planetary Laboratory, University of Arizona, Tucson, AZ, 85721, USA}

\author[5,6]{Sarah Peacock}
\affil[5]{University of Maryland, Baltimore County, MD 21250, USA}
\affil[6]{NASA Goddard Space Flight Center, Greenbelt, MD 20771, USA}

\author[1]{Kevin B. Stevenson}

\author[3]{Mercedes L\'opez-Morales}

\author[7,8]{Ryan J. MacDonald}
\affil[7]{Department of Astronomy, University of Michigan, Ann Arbor, MI, USA}
\affil[8]{NHFP Sagan Fellow}

\author[1]{L. C. Mayorga}

\author[2,9]{David K. Sing}
\affil[9]{Department of Earth \& Planetary Sciences, Johns Hopkins University, Baltimore, MD, USA}

\author[1]{Kristin S. Sotzen}

\author[10]{Jeff A. Valenti}
\affil[10]{Space Telescope Science Institute, Baltimore, MD 21218, USA}

\author[3]{Jea Adams}

\author[11]{Munazza K. Alam}
\affil[11]{Carnegie Earth \& Planets Laboratory, Washington, DC, 20015, USA}

\author[12]{Natasha E. Batalha}
\affil[12]{NASA Ames Research Center, Moffett Field, CA, USA}

\author[9]{Katherine A. Bennett}

\author[9]{Junellie Gonzalez-Quiles}

\author[13]{James Kirk}
\affil[13]{Department of Physics, Imperial College London, Prince Consort Road, London, SW7 2AZ, UK}

\author[5,6]{Ethan Kruse}

\author[14]{Joshua D. Lothringer}
\affil[14]{Department of Physics, Utah Valley University, Orem, UT, 84058 USA}

\author[9]{Zafar Rustamkulov}

\author[15]{Hannah R. Wakeford}
\affil[15]{School of Physics, HH Wills Physics Laboratory, University of Bristol, Bristol, UK}




\abstract{
The critical first step in the search for life on exoplanets over the next decade \cite{Fujii2018, ExoScienceStrategy2018} is to determine whether rocky planets transiting small M-dwarf stars possess atmospheres \cite{Seager2009, Mansfield2019} and, if so, what processes sculpt them over time \cite{Segura2005, Tian2015, Luger2015}. Because of its broad wavelength coverage and improved resolution compared to previous methods, spectroscopy with JWST offers a new capability to detect and characterize the atmospheres of Earth-sized, M-dwarf planets \cite{Batalha2018, Lustig-Yaeger2019}.  
Here we use JWST to independently validate the discovery of LHS~475b \citep{Guerrero2021}, a warm (586 K), 0.99 Earth-radius exoplanet, interior to the habitable zone, and report a precise 2.9 $-$ 5.3 {\micron} transmission spectrum.  With two transit observations, we rule out primordial hydrogen-dominated and cloudless pure methane atmospheres.  Thus far, the featureless transmission spectrum remains consistent with a planet that has a high-altitude cloud deck (similar to Venus), a tenuous atmosphere (similar to Mars), or no appreciable atmosphere at all (akin to Mercury).  There are no signs of stellar contamination due to spots or faculae \citep{Rackham2018}. Our observations demonstrate that JWST has the requisite sensitivity to constrain the secondary atmospheres of terrestrial exoplanets with absorption features ${<}$ 50 ppm, and that our current atmospheric constraints speak to the nature of the planet itself, rather than instrumental limits. 
}

\keywords{JWST, Terrestrial Exoplanet Atmospheres, Transmission Spectroscopy}

\maketitle

\clearpage

The search for atmospheres on rocky exoplanets has only just begun. Prior constraints on the presence of terrestrial exoplanet atmospheres using the Hubble Space Telescope (HST) and the Spitzer Space Telescope (Spitzer) have succeeded in ruling out primordial \ce{H2}/He atmospheres \citep{Delrez2018, deWit2016, deWit2018, Zhang2018, Edwards2021} that would produce large and detectable absorption features in a transmission spectrum; thick atmospheres that would produce shallow infrared secondary eclipse depths \citep{Kreidberg2019}; and a tentative detection of a terrestrial atmosphere \citep{Swain2021} that remains controversial \citep{Mugnai2021, LibbyRoberts2022}. JWST is expected to break new ground in the search for atmospheres on rocky exoplanets that transit nearby M dwarfs \citep{Morley2017, Lustig-Yaeger2019}. However, theoretical modeling work predicts a tumultuous stellar environment in these compact M-dwarf systems \citep{Garcia-Sage2017, Airapetian2020} that raises the critical question of whether or not small, rocky exoplanets can maintain thick and detectable atmospheres in the face of significant atmospheric loss processes.  

We observed two transits of LHS~475b (previously the planet candidate TOI 910.01) on 31 August 2022 and 4 September 2022 with JWST's Near InfraRed Spectrograph (NIRSpec) \cite{NIRSpec2022, NIRSpec2022_Exoplanets} G395H instrument mode as part of the JWST Cycle 1 Guest Observing (GO) Program 1981 (PI: K. Stevenson). This mode covers wavelengths $2.87-5.27$ {\microns} and has a native spectral resolving power of $\mathcal{R} = \lambda / \Delta \lambda \approx 2700$.  We used the Bright Object Time Series (BOTS) mode with the NRSRAPID readout pattern, S1600A1 slit, and the SUB2048 subarray. Each time-series observation lasted a total of 2.9 hours, which captured the $39.98 \pm 4.04$ minute transit that occurred during this window. This resulted in approximately 1.75 hours and 0.5 hours of stellar baseline before and after transit, respectively.  The Methods contains additional information on the observations.

We selected the LHS~475 system as one of several nearby M-dwarf systems with known or candidate rocky planets.  Prior to validation, LHS~475b was classified as a planet candidate first identified in Sector 12 by the Transiting Exoplanet Survey Satellite (TESS) \cite{Ricker2015}.  TESS observed subsequent transits of the planet in Sectors 13, 27, and 39.  LHS~475b transits a $3300$ K, $0.2789$ R$_{\odot}$ M3.5V dwarf star on a 2.029-day orbital period \cite{Stassun2019}. This planet is likely to be tidally locked, with a permanent dayside facing its host star \citep{Kasting1993}, and an equilibrium temperature of $586$ K. 

We validated the discovery of LHS~475b by eliminating both instrumental and astrophysical false positives.  JWST detected two transit signals at the predicted times that are consistent in depth and duration with the 45 TESS transits ($978 \pm 73$ ppm, $42 \pm 13$ minutes). Archival DSS images from 1999 rule out the possibility of a background transiting star-planet system or an eclipsing binary.  LHS~475 is a high-proper-motion star (1.28 arcsec/year) and no flux sources were identified in the archival data along its path from 1999 to 2022. See the Methods for more details about the archival imagery. In parallel with this work, Ment et al. (in prep) performed an independent validation of \planetname using ground-based follow-up observations.

\begin{figure}[t] 
    \centering
    \includegraphics[width=\textwidth]{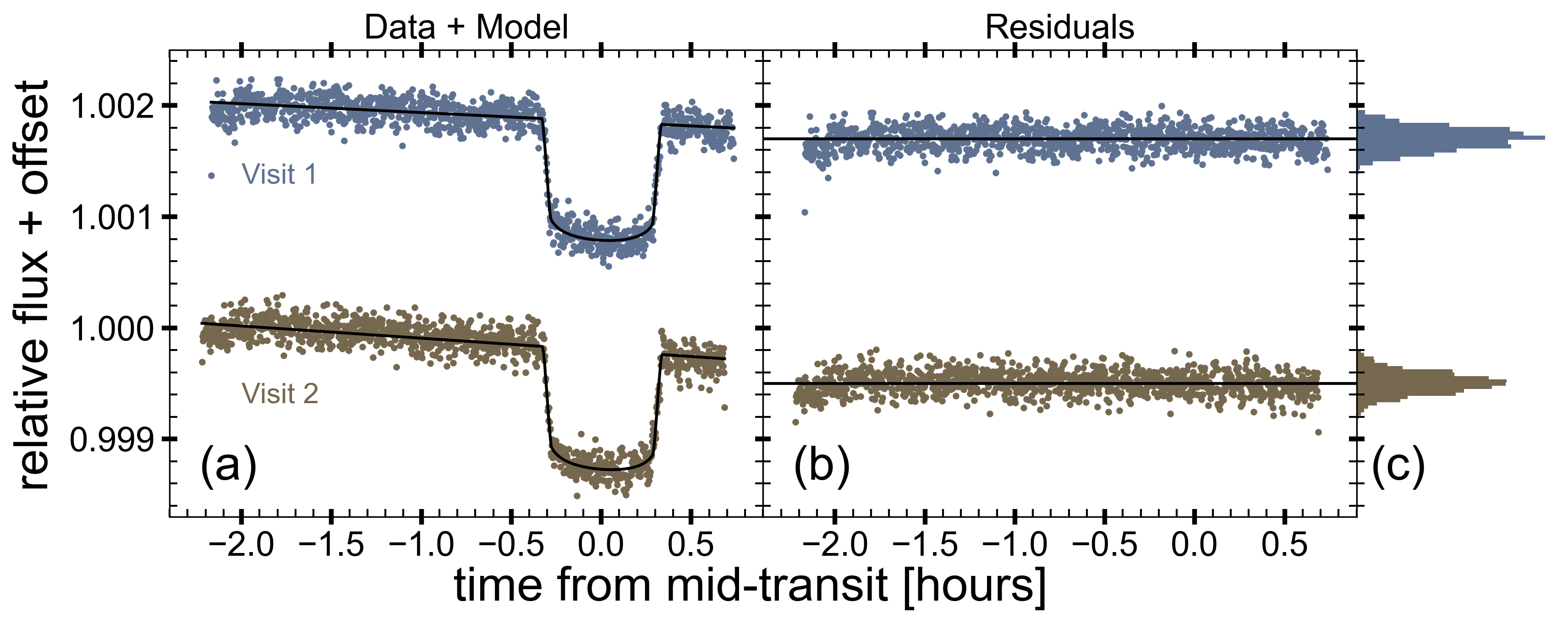}
    \caption{\label{fig:WLC}{White light curves from both LHS~475b visits using the \firefly reduction (see Methods). For each visit, we combined data from both the NRS1 and NRS2 detectors into a single white light curve and applied a vertical offset for clarity. A transit model and visit-long linear trend are sufficient to fit the raw white light curves (panel a).  Residuals from the best fit (panel b) highlight a small, $\sim$15-minute ramp at the start of each visit. Residuals are shown on the same y-scale as panel a. Both histograms of the residuals (panel c) are Gaussian distributed. 
    }} 
\end{figure}

We reduced the JWST data using three independent pipelines---\eureka \citep{Eureka2022}, \firefly \citep{Rustamkulov2022b}, and Tiberius \citep{2018MNRAS.474..876K, 2019AJ....158..144K, 2021AJ....162...34K}---that yielded consistent results (within $1.1\sigma$, see Methods for details on each analysis).  For our final interpretation, we utilize results from the \firefly pipeline as it is the most representative of the three reductions.  We generated white light curves across the full G395H wavelength range covered by the two detectors, NRS1 from 2.884 - 3.720 {\microns} and NRS2 from 3.820 - 5.177 {\microns}. The planet transits are clearly visible in the raw white light curves (see \Cref{fig:WLC}).  We note the presence of a small ramp at the start of the observation; no additional structure is seen in the residuals. There is also no evidence of starspot crossings during the transits. Our joint fit to the white light curves gives a planet-to-star radius ratio of $R_p/R_s = 0.03257 \pm 0.00014$, a mid-transit time of $T_0 = 59822.8762593 \pm 0.000026$ BMJD$_{\rm TDB}$, an orbital period of $P = 2.029088 \pm 0.000006$ days, a ratio of the semi-major axis to the stellar radius of $a/R_s = 15.87235 \pm 0.472$, and an inclination of $i = 87.194^{\circ} \pm 1.39^{\circ}$. Therefore, \planetname has a radius of $\mathrm{R}_p = 0.99 \pm 0.05~\mathrm{R}_{\oplus}$ ($6319 \pm 318$ km). 
Although \planetname's mass has not been measured, assuming an interior composition that is consistent with the small, rocky M-dwarf exoplanet population \citep{Luque2022}, we estimate a planet mass of $\mathrm{M}_p = 0.914 \pm 0.187 ~\mathrm{M}_{\oplus}$. 

\begin{figure}[t]
    \centering
    \includegraphics[width=0.99\linewidth]{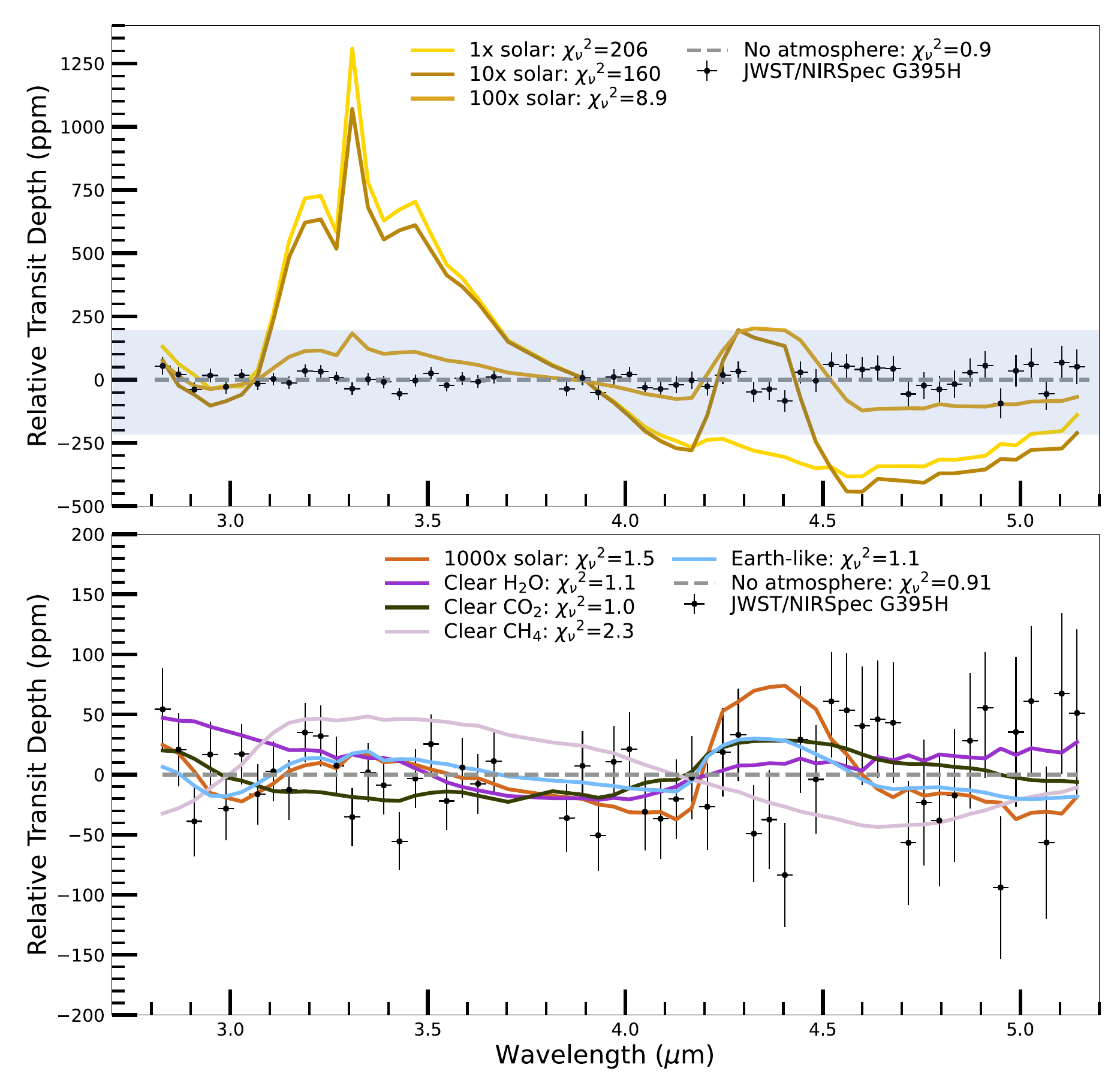}
    \caption{\label{fig:spectrum_zoom}{
    Final, binned spectrum (black points) compared to models (coloured lines). Top: Our data strongly ($\gt$ 10$\sigma$) rule out hydrogen-dominated atmospheres with compositions from 1$\times$ -- 100$\times$ solar metallicity, with reduced-$\chi^2$s reported in the legend for each model. The blue shaded bar highlights the region detailed in the bottom panel. Bottom: Our data also rule out, though to lower (2--5$\sigma$) significance, high mean molecular weight compositions of 1000$\times$ solar metallicity or a pure methane atmosphere. We weakly disfavor a pure water atmosphere or an Earth composition atmosphere. The data are consistent with a pure carbon dioxide atmosphere or that of an airless body. Each model is plotted relative to the mean transit depth. 
    }}
\end{figure}

We fitted the spectroscopic light curves at the detector's pixel resolution to derive wavelength-dependent transit depths independently for the first and second visit. The orbital parameters were fixed to the values from the joint white light curve analysis, leaving the planet-to-star radius ratio, linear temporal slope, and a constant offset as free parameters. We adopted stellar limb darkening from a 3D stellar model grid \citep{Magic2015}. We then performed a weighted average of spectra from the two visits and binned the combined native-pixel-resolution transit depths into 56 points ($\mathcal{R} \approx 100$).
Our co-added and binned transmission spectrum is shown in \Cref{fig:spectrum_zoom}.  

The observed transmission spectrum is featureless. A flat line, representative of an airless-body or high mean molecular weight (MMW) atmosphere, fitted to the binned data produces a reduced $\chi^2 = 0.91$. No evidence is seen for stellar contamination from unocculted cool spots or hot faculae on the stellar disk \citep[e.g.,][see Methods]{Rackham2018, Zhang2018}. Despite the featureless spectrum, the precision is sufficiently high to rule out ($\gt 5\sigma$) several archetypal atmospheric compositions, including primordial hydrogen-helium atmospheres with less than 100 ${\times}$ solar metallicity, as well as pure \ce{CH4} atmospheres $\ge 1$ bar. We can specifically rule out this pure \ce{CH4} atmospheres due to the low mass of the \ce{CH4} molecule and the presence of the strong 3.3 \micron{} \ce{CH4} band in the G395H bandpass. Other secondary atmospheres are more challenging to rule out and distinguish from one another, however. We only weakly disfavor (at $\gtrsim$1$\sigma$; \citep{Trotta2008}) 1000 ${\times}$ solar metallicity, pure steam, or warm Earth-like atmospheric compositions.  Both a pure $\ge 1$ bar carbon dioxide atmosphere or no atmosphere are favored yet are statistically indistinguishable from each other. In the context of Solar System terrestrial archetype atmospheres (see Methods Fig. \ref{fig:solar_system_models}), we also weakly disfavor ($\gtrsim$1$\sigma$) clear, warm Venus-like and Titan-like atmospheres. We cannot statistically distinguish between a thin Mars-like atmosphere, a hazy Titan-like atmosphere, and a cloudy Venus, which are all consistent with the data. 

\begin{figure}[t]
    \centering
    \includegraphics[width = \textwidth]{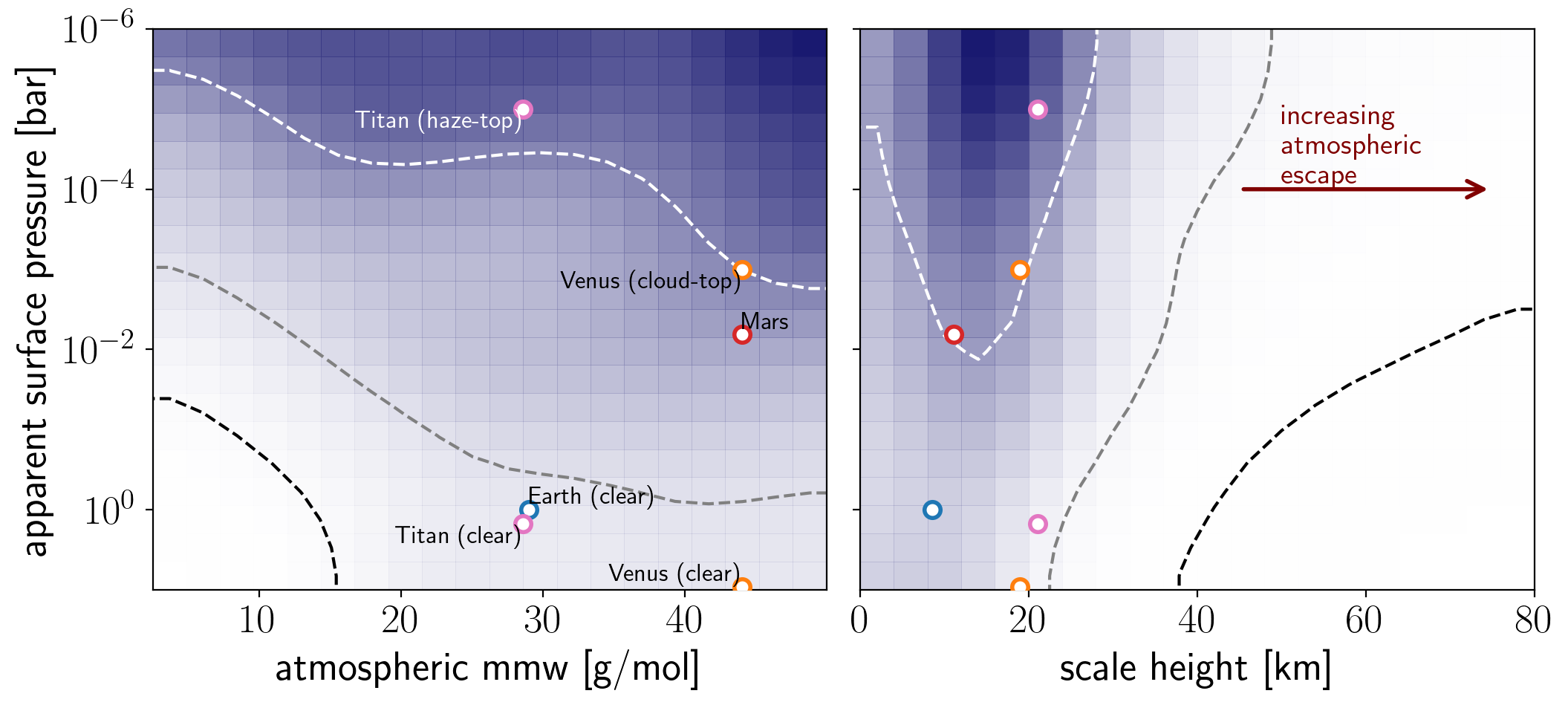}
    \caption{Retrieval results showing preferred atmospheric properties for models containing \ce{H2O}, \ce{CO2}, \ce{CH4}, and \ce{CO}, plus a variable bulk gas composition for LHS~475b given the transmission spectrum measurements. Darker color shading indicates higher relative posterior probability density as a function of the apparent surface pressure ($P_0$), molecular weight ($\mu$) of the bulk atmospheric composition (left), and isothermal scale height ($H$, right). Dashed contours denote the $1 \sigma$ (white), $2 \sigma$ (gray), and $3 \sigma$ (black) Bayesian credible regions. The red arrow depicts how the Jeans escape flux depends on the scale height and emphasizes the region of the parameter space that is more susceptible to atmospheric escape.   
    If the planet possesses an atmosphere with at least 1 ppm \ce{CO2} or \ce{CH4}, then the models prefer high mean molecular weight, compact atmospheres ($\mu > 20$ g/mol at $1.2 \sigma$; $H < 25$ km at $1.2 \sigma$) with low apparent surface pressures ($P_0 < 0.01$ bar at $1.3 \sigma$; $P_0 < 1$ bar at $2 \sigma$). These scenarios correspond to either a tenuous or cloudy secondary atmosphere. 
    }
    \label{fig:retrieval_pair}
\end{figure}

Following previous analyses \citep{Kreidberg2014}, we performed Bayesian retrievals to better explore the range of atmospheres that remain consistent with our spectroscopic measurements. We assumed a five component atmospheric composition consisting of the four most common and spectroscopically active molecules (\ce{H2O}, \ce{CO2}, \ce{CH4}, and \ce{CO}) in the Solar System terrestrial atmospheres, plus an unspecified gas that constitutes the bulk atmospheric composition but is spectroscopically inactive at these wavelengths \citep[e.g.][]{Tremblay2020}. We allow the mean molecular weight of the bulk gas to vary between 2.5 g/mol and 50 g/mol. Since a solid planetary surface and an optically thick gray cloud deck are indistinguishable in the transit spectrum, we fit for the apparent surface pressure (the pressure of an opaque surface above which the atmosphere extends). We marginalize over the aforementioned planet mass, which is assumed to be consistent with a rocky interior composition. The vertical extent of the atmosphere is dictated by the scale height, which is implicitly controlled by varying the atmospheric temperature, mean molecular weight, and planet gravity. See Figure \ref{fig:retrieval_pair} for the retrieval results summarizing the range of allowed atmospheres given our data and highlighting the degeneracies that persist among the remaining atmospheric possibilities.  

If the planet has an atmosphere, it is likely to be a high mean molecular weight secondary atmosphere that is tenuous (Mars-like) or cloudy/hazy (Venus-like or Titan-like). Compact atmospheres with small scale heights are preferred across the full range of apparent surface pressures. High mean molecular weight atmospheres dominated by species heavier than $40$ g/mol, like \ce{CO2} or Argon, can be thicker while maintaining relatively flat spectra. Atmospheric characteristics that increase the scale height, including high temperatures and low mean molecular weight bulk atmospheric compositions, tend to be disfavored, particularly for apparent surface pressures ${\gtrsim} 10$ mbar. Models with scale heights larger than 20 km strongly skew towards the low mean molecular weight atmospheres ($\mu < 10$ g/mol), make up ${\sim} 50 \%$ of the lowest apparent surface pressure samples, and tend to have low abundances of all absorbing molecules. These extended atmospheres with low apparent surface pressures are unlikely to form clouds or hazes at such high altitudes and are the most susceptible to atmospheric loss, making them less physically plausible scenarios. 
Although LHS 475 is typical of low-activity M dwarfs in the solar neighborhood \citep{Medina2020}, atmospheric escape processes are still a concern for a primordial extended atmosphere, and if LHS 475 b is indeed airless, such processes would likely constitute the primary reason for this. 

Our two transit observations demonstrate that JWST has the sensitivity to detect and constrain the secondary atmospheres of terrestrial exoplanets, and therefore our atmospheric non-detection reflects the nature of the target itself. We place a $3\sigma$ constraint on the maximum size of absorption features in our spectrum at 61 ppm for \ce{H2O} at 2.8 \micron{}, 38 ppm for \ce{CH4} at 3.3 \micron{}, 49 ppm for \ce{CO2} at 4.3 \micron{}, and 62 ppm for \ce{CO} at 4.6 \micron{}.  These constraints demonstrate JWST's sensitivity to absorption features smaller than 50 ppm for an Earth-sized exoplanet. We find no indication of a noise floor down to 5 ppm (See Methods \Cref{fig:allan}). These are critical benchmarks for forthcoming rocky exoplanet observations with JWST. Furthermore, our non-detection of starspot crossings during transit and the lack of stellar contamination in the transmission spectrum are promising signs in this initial reconnaissance of \planetname. These findings indicate that additional transit observations of \planetname with JWST are likely to tighten the constraints on a possible atmosphere. A third transit of \planetname is scheduled as part of this program (GO 1981) in 2023. An alternative path to break the degeneracy between a cloudy planet and an airless body is to obtain thermal emission measurements of \planetname during secondary eclipse because an airless body is expected to be several hundred Kelvin hotter than a cloudy world and will therefore produce large and detectable eclipse depths at JWST's MIRI wavelengths \citep{Koll2019, Mansfield2019}. Our findings only skim the surface of what is possible with JWST. 

\subsection*{Acknowledgements}

This work is based in part on observations made with the NASA/ESA/CSA JWST. The data were obtained from the Mikulski Archive for Space Telescopes at the Space Telescope Science Institute, which is operated by the Association of Universities for Research in Astronomy, Inc., under NASA contract NAS 5-03127 for JWST.  These observations are associated with program \#1981.
Support for program \#1981 was provided by NASA through a grant from the Space Telescope Science Institute, which is operated by the Association of Universities for Research in Astronomy, Inc., under NASA contract NAS 5-03127.

\newpage
\backmatter

\bmhead{Methods}

\section{Data Analysis}

\begin{figure}[t] 
    \centering
    \includegraphics[width=\textwidth]{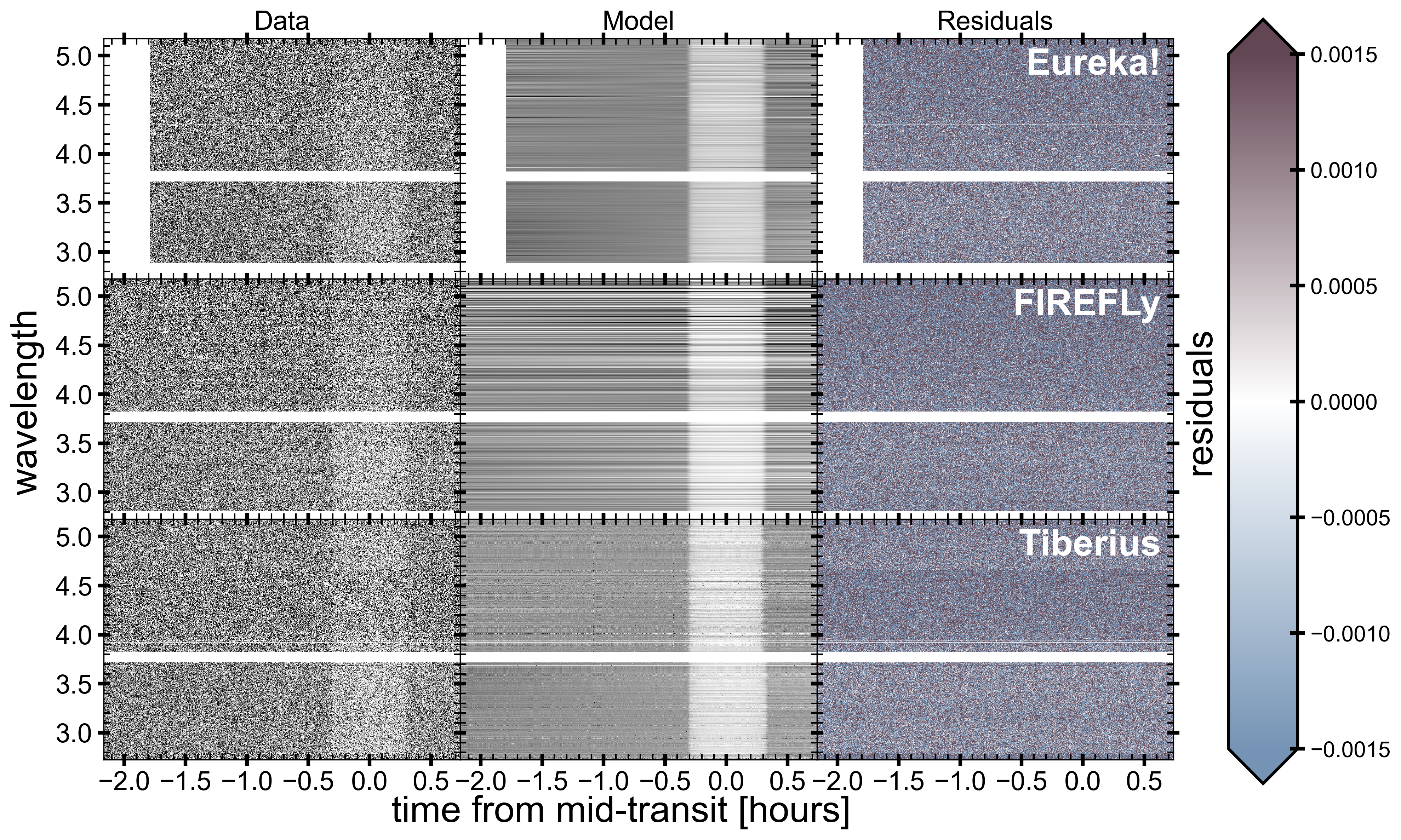}
    \caption{\label{fig:2dLC}{
    2D light curves of LHS~475b as a function of time and wavelength for the first visit, measured with NIRSpec/G395H. The horizontal stripe down the middle of each panel corresponds to the gap between the NRS1 and NRS2 detectors. Left: data normalized by the median stellar spectrum. Middle: Maximum probability transit models. Right: Residuals from the model fit. Note, the \eureka and \firefly reductions trim more blue columns from NRS1 where there is minimal throughput than the Tiberius pipeline, which accounts for the regions without data in those reductions. Similarly, the \eureka reduction also trims off the initial ramp that can be seen in Figure \ref{fig:WLC}.
    }} 
\end{figure}

\subsection{Observations}

We observed two transits of LHS~475b with the NIRSpec G395H grating covering the 2.87–5.14 \micron{} wavelength range split over the NRS1 and NRS2 detectors, with a detector gap between 3.72 and 3.82 \micron{}. The first transit was observed on the 31 August 2022 18:48 UTC and the second on 4 September 2022 20:09 UTC. Each visit lasted 4.4 hours in total with 2.9 hours of exposure time. Both transits were executed with the same observing settings, using the Bright Object Time Series (BOTS) mode with the NRSRAPID readout pattern, S1600A1 slit, and the SUB2048 subarray from NIRSpec. We obtained a total of 1158 integrations per visit, with 9 groups per integration and 0.902 seconds per group.

We extracted and analyzed the data from each visit independently with the \eureka, \firefly and Tiberius pipelines as described below. 2D lightcurves, models, and residuals from the three reductions are shown in Figure \ref{fig:2dLC}. 

\subsection{Spectral Extraction}

\subsubsection{Eureka!}

\eureka \citep{Eureka2022} is an end-to-end analysis pipeline for time series observations (TSOs) of exoplanets. \eureka\ serves as a wrapper for stages 1 and 2 of the \texttt{jwst} pipeline \citep{jwstpipeline2022}, allowing the user to specify which steps are run in addition to custom modules. In later stages, \eureka\ performs spectroscopic extraction, light curve generation, and light curve fitting.

In this work, we apply \eureka's custom group-level background subtraction (GLBS) in stage 1 prior to ramp fitting to remove 1/f noise which has been found to impact the accuracy of ramp fits for data with a small numbers of groups up the ramp \citep{ERSFirstLook, Rustamkulov2022b}. Due to G395H's curved trace we first identify the center of the trace, then mask all pixels within an aperture of 8 pixels. All remaining pixels in a given column (cross-dispersion direction) were used to calculate a median background/noise level for that column. We skip the jump step detection, otherwise running all standard stage 1 steps for TSOs. 

In stage 2, we skip the flat field step (at the time of writing, only pre-flight flat fields were available, which are insufficient for the precision we require and adds significant noise to the data) and the photom step. Because we are interested in relative flux measurements, we do not require the absolute flux calibration provided by these two steps.

A second round of background subtraction is done in stage 3 to capture any remaining background or 1/f noise, using pixels more than 9 pixels away from the center of the trace. The spectrum is extracted with an aperture of 5 pixels for NRS1 and 4 pixels for NRS2 using median frame optimal spectral extraction. To convert from DN/s to electrons, we apply a median of the gain files. At the time of writing only pre-flight gain files were available, which are insufficient for the precision we require and adds significant noise to the data if applied on a per-pixel basis. For NRS1 we extract only columns $800-2047$ due to the negligible throughput outside of that region of the detector. For NRS2 we extract the full dispersion direction, but note that the edges are less reliable due to the trace approaching the top or bottom of the subarray. 

\begin{figure}
    \centering
    \includegraphics[width = \textwidth]{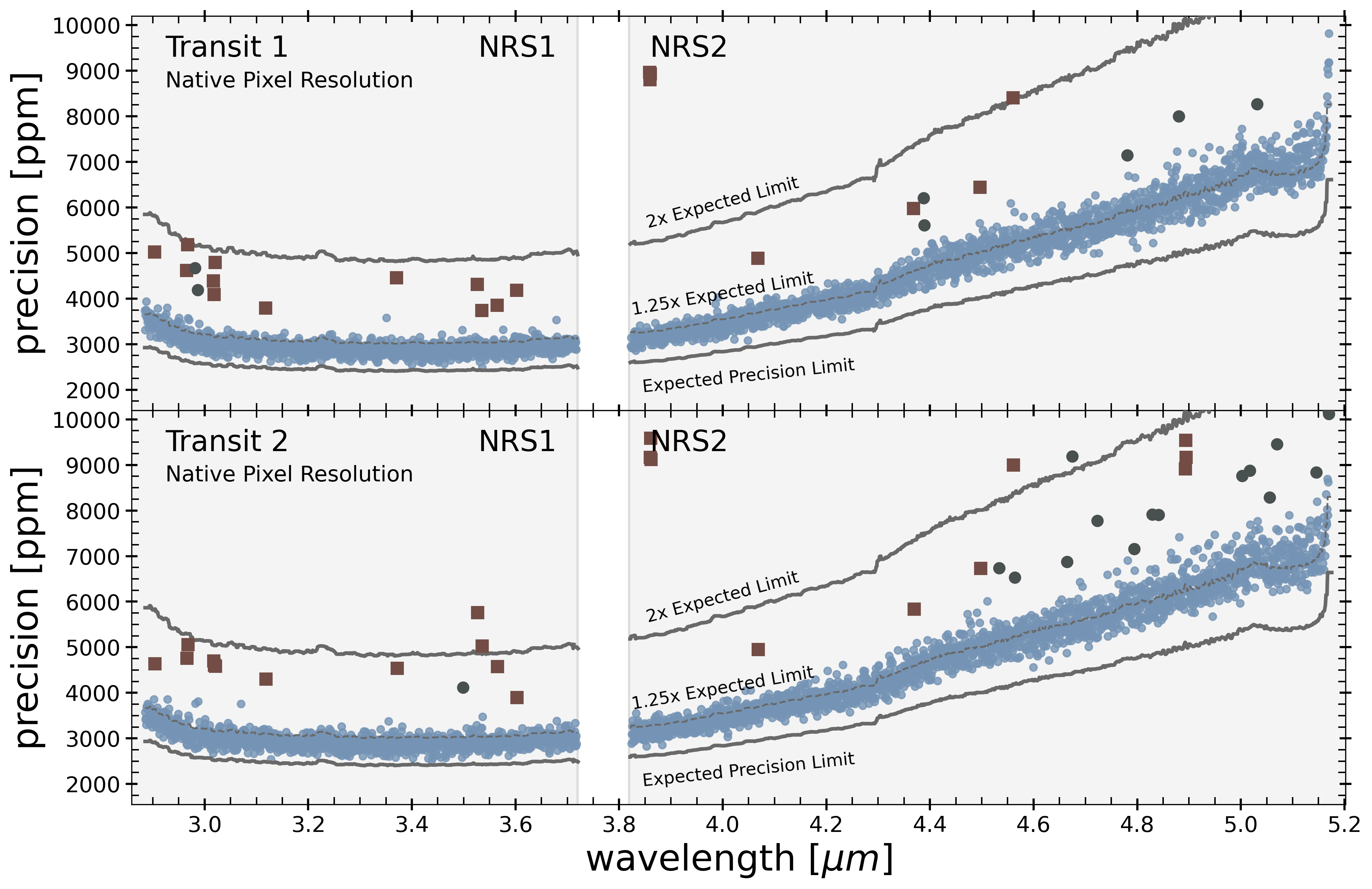}
    \caption{\eureka\ spectrophotometric precision at the native pixel resolution, compared to expected noise levels for both events. The expected noise level, as well as 1.25$\times$ and 2$\times$ the expected noise are shown as grey lines, these have been smoothed to the resolution of the final transit spectrum for visualization purposes. Squares denote columns which are greater than 1.5$\times$ the expected noise level in both transits, dark blue circles denote columns which are greater than 1.5$\times$ the expected noise level in only one transit. These columns are flagged and not used to generate the final transmission spectrum.}
    \label{fig:eurka_spec_precision}
\end{figure}

White light curves are generated across the full wavelength range of the extracted data: 2.884 - 3.720 \micron{} for NRS1 and 3.820 - 5.177 \micron{} for NRS2. For transit 1 we reach a white light precision of 112 ppm and 162 ppm for NRS1 and NRS2, respectively. For transit 2 we reach a white light precision of 116 ppm and 149 ppm for NRS1 and NRS2, respectively. For each transit, NRS1 and NRS2 are combined into a single white light curve prior to light curve fitting. We extract spectroscopic light curves at the pixel-resolution following recommendations from \cite{Espinoza2022}, however we find that our GLBS routine sufficiently removes the 1/f noise in our data set, with no improvement on the final transmission spectrum precision between fitting light curves at the native pixel resolution and then binning, or binning prior to fitting (see Section \ref{sec:eureka_lcfit}). Figure \ref{fig:eurka_spec_precision} shows our spectroscopic precision compared to expected noise levels. Bad columns are denoted by squares (flagged in both transits) or darker circles (flagged in only one transit). This corresponds to 1.13\% and 1.56\% of columns in transit 1 NRS1 and NRS2, respectively, and 1.05\% and 2.10\% of columns in transit 2 NRS1 and NRS2, respectively. Excluding these columns, for transit 1 we achieve a median precision of 1.19$\times$ and 1.23$\times$ the expected noise level for NRS1 and NRS2, respectively, while for transit 2 we achieve 1.19$\times$ and 1.24$\times$ the expected noise level for NRS1 and NRS2, respectively.

\subsubsection{FIREFLy} 
We used the \firefly \citep[Fast InfraRed Exoplanet Fitting for Lightcurves,][]{Rustamkulov2022,Rustamkulov2022b} to analyse the JWST data.
We started with the \texttt{uncal.fits} files and ran the \texttt{jwst} pipeline for stage 1 and 2 with modified steps including group-level 1/f and background subtraction and skipping the jump-step. Both changes were shown to decrease the scattering in the extracted lightcurves. After obtaining the \texttt{rateints.fits} files from the stage 2 output, we performed custom cosmic rays and bad/hot pixels corrections. The spectral traces are then masked in the cleaned 2D images before applying 1/f correction, which subtracts the median value of the unmasked background pixels at each column. Next, we measured the shifts of the spectral trace in x and y directions by using cross correlation in the selected 2D spectral region. The measured shifts are less than one hundredth of a pixel which illustrates the excellent pointing stability of JWST. After aligning each 2D spectrum, we determine the spectral trace by first cross correlating a Gaussian profile at each column to obtain the spectrum location in the y direction, and then fit a 4th order polynomial as a function of the x direction. The spectrum is then extracted for each integration centered at the fitted spectral trace to form the light curves. 

\subsubsection{Tiberius}

The Tiberius pipeline is a spectral extraction and light-curve fitting code based on the LRG-BEASTS pipeline \citep{2018MNRAS.474..876K, 2019AJ....158..144K, 2021AJ....162...34K}. We used Tiberius on the Eureka! stage 1 group-level background-subtracted product, which had 1/f noise removed, to produce white and spectroscopic transit light curves. First we created bad-pixel masks for NRS1 and NRS2 by manually selecting hot pixels in the data. These hot pixels were combined with all pixels flagged as 3$\sigma$ outliers from the background, and were interpolated over using their nearest neighboring pixels. We also interpolate the spatial dimension of the data on a 10x grid, which improves flux extraction at the sub-pixel level, reducing noise. The spectra were then traced by fitting Gaussian functions for each column of the detectors, and then using a running median to smooth the trace centers. These centers were fit with a 4th-order polynomial, 3$\sigma$ outliers were removed, and the centers were again refit with a 4th-order polynomial. 

In addition to the background subtraction already performed in the creation of the stage 1 product, we perform an additional background subtraction step here to remove residual background light or remaining 1/f noise. We mask from the detector a defined aperture of 4 pixels plus 6 more pixels offset from it, and clipped 3$\sigma$ outliers in the background pixels, with respect to their specific column and frame. Finally, the background signal for each column was subtracted from it, and the spectra were then extracted using a 4 pixel aperture. 

\subsection{Light Curve Fitting}

\subsubsection{Eureka!} \label{sec:eureka_lcfit}
We perform a joint fit on both white light curves to constrain the system parameters. Limb darkening is calculated with the \texttt{ExoTic-LD} pacakge \citep{Laginja2020, iva_laginja_2020_3923986, hannah_wakeford_2022_6809899} using a quadratic limb darkening \citep{claret2000, sing2010} and the 3D stellar grid from \cite{Magic2015}. Stellar parameters are adopted from \citep{Magic2015}, assuming T$_{eff}$ = 3312 K, log(g) = 4.94, Fe/H = 0.0. For all light curve fits both limb darkening parameters are held constant.  We find that the uncertainty induced in the light curve fits by the stellar models is smaller than the uncertainty in individual transit depths, with consistent transit spectra regardless of free or fixed limb darkening. We trim the first 150 integrations prior to light curve fitting to remove a slight ramp at the beginning of the data, which can be seen in Figure \ref{fig:WLC}.

For all light curve fitting we consider a transit model \citep[\texttt{batman},][]{batman2015} and a linear ramp in time. We use \texttt{emcee} \citep{emcee2013}, running each chain to at least 10$\times$ the auto-correlation time. The joint white light curve fit includes the planet radius, orbital period, center of transit, inclination, and scaled semi-major axis as shared parameters, and independent temporal ramps for each white light curve. Best fit orbital parameters are given in Table \ref{tab:orbital_parameters}. The \eureka spectroscopic fits adopt the \eureka white light best-fit orbital parameters and only fit for planet radius and the linear temporal ramp. Following \cite{Espinoza2022} we extract and fit our light curves at the native pixel resolution of the detectors, and later bin the data to our preferred resolution. 

\begin{figure}
    \centering
    \includegraphics[width = \textwidth]{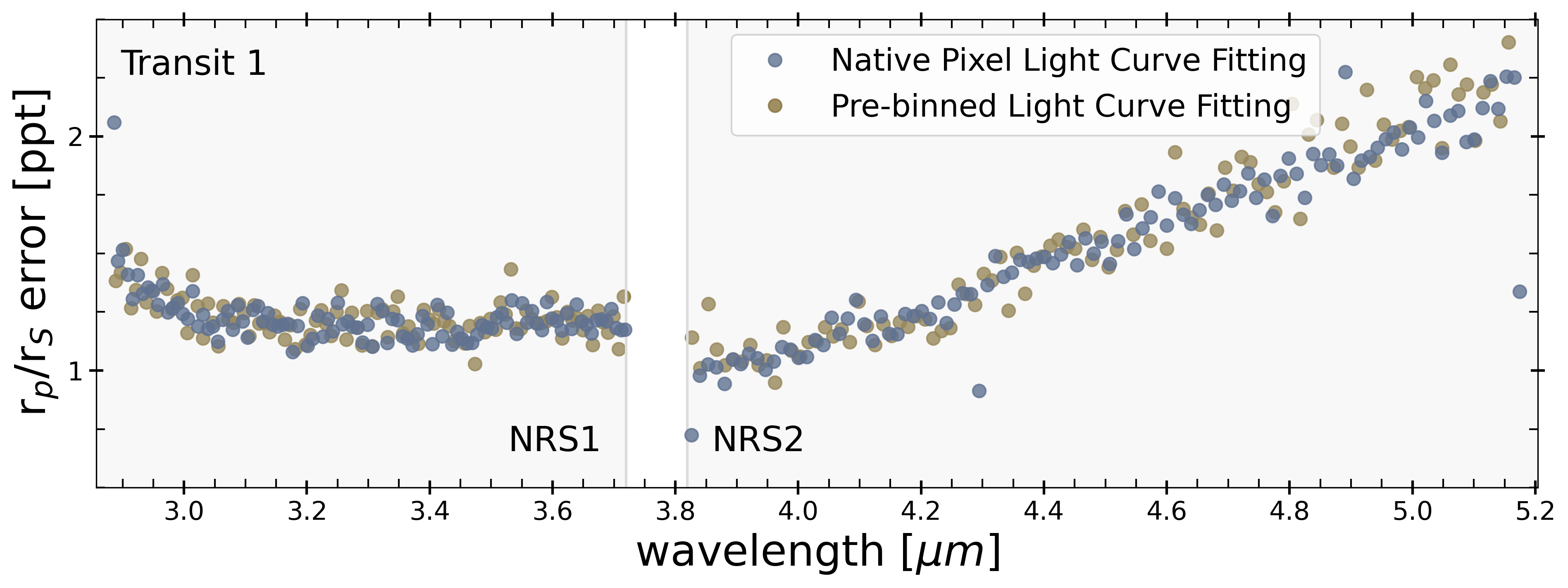}
    \caption{Comparison of uncertainty on planet radius derived from light curves fit at the native pixel resolution and fitting of pre-binned light curves. The y-axis is in parts per thousand. We find little to no difference in the uncertainty, suggesting that our 1/f correction is sufficient to address the column-column variances.}
    \label{fig:error_comp}
\end{figure}

To test the robustness of our group-level 1/f noise correction, we also fit a set of pre-binned light curves and compare the resulting uncertainty on the planet radius. Figure \ref{fig:error_comp} shows our uncertainty on planet radius for transit 1 for both the native pixel resolution light curve fitting, and our pre-binned light curve fitting. We find no significant improvement by fitting the full resolution light curves, suggesting the 1/f noise has been sufficiently removed. We suggest that this test should be run on all NIRSpec G395H TSOs to ensure that one has sufficiently removed the 1/f noise.

\subsubsection{FIREFLy}

The extracted light curves are first summed in the wavelength direction including both NRS1 and NRS2 to form the whitelight light curve for each visit. We then used \texttt{batman} \citep{batman2015} and \texttt{emcee} \citep{emcee2013} to joint fit the whitelight lightcurves from the two visits with six free parameters including R$_{planet}$/R$_{star}$, a/R$_{star}$, orbital inclination, mid-transit time for both visits, and linear temporal slope for both visits. The best-fit joint white light orbital parameters are listed in Table \ref{tab:orbital_parameters}. We fixed the limb darkening to the quadratic coefficients from the 3D stellar model in the Stagger-grid \citep{Magic2015} interpolated at Fe/H=0, T$_{eff}$=3312K and log(g)=4.94.

The orbital parameters and quadratic limb darkening coefficients are then fixed to fit the lightcurve from each wavelength column. We used the \texttt{scipy.optimize.curvefit} function with three free parameters including linear temporal slope, constant offset and R$_{planet}$/R$_{star}$. 

\subsubsection{Tiberius}

We extracted a white light curve for each detector (NRS1 and NRS2), for each transit. These white light curves were fit independently using a Levenberg–Marquardt damped least squares routine with the Tiberius pipeline. Limb darkening parameters were obtained with LDTK \cite{Parviainen2015, Husser2013} from assumed stellar parameters of T$_{eff}$ = 3312 K, log(g) = 4.94, Fe/H = 0.0, and a quadratic limb darkening law was used. The results of the white light curve fits were used to fix the transit parameters for the spectroscopic light curve fits, which were performed at pixel-level resolution using the same damped least squares routine. Since the white light curves were fit independently, the best-fit parameters in Table \ref{tab:orbital_parameters} were obtained from a weighted average of the results of each of the four white light curve fits (weighted by flux received on each detector).

\begin{table}[]
    \centering
    \tiny
    \setlength\tabcolsep{2.5pt}
    {\def\arraystretch{2}
    \begin{tabular}{|r l || r l | r l | r l |}
    \hline
        \multicolumn{2}{|c||}{Parameter}  & \multicolumn{2}{c|}{\eureka} & \multicolumn{2}{c|}{\firefly} & \multicolumn{2}{c|}{Tiberius} \\ \hline \hline
        R$_p$/R$_s$ & [unitless]                & 0.032756 & $^{+1.44\times10^{-4}}_{-1.44\times10^{-4}}$      &  0.03257 & $^{+1.40\times10^{-4}}_{-1.43\times10^{-4}}$  & 0.032226  &  $^{+4.86\times10^{-4}}_{-4.86\times10^{-4}}$     \\ 
        T$_0$       & [BMJD$_{TDB}$]    & 59822.8762805 & $^{+2.92\times10^{-5}}_{-2.91\times10^{-5}}$      &  59822.8762593 & $^{+2.62\times10^{-5}}_{-2.62\times10^{-5}}$   &  59822.8763396 &  $^{+5.85\times10^{-5}}_{-5.85\times10^{-5}}$ \\
        Period      & [days]            & 2.02908843  & $^{+5.65\times10^{-6}}_{-5.66\times10^{-6}}$   &  2.0290882  & (Fixed)  &  2.02909  &  (Fixed) \\
        a/R$_s$     & [unitless]                & 15.223  & $^{+4.62\times10^{-1}}_{-4.37\times10^{-1}}$    &  15.87235  & $^{+4.88\times10^{-1}}_{-4.56\times10^{-1}}$  &  18.161  & $^{+1.79}_{-1.79}$  \\
        i           & [degrees]         & 86.991  & $^{+1.41\times10^{-1}}_{-1.39\times10^{-1}}$    &  87.194  & $^{+1.41\times10^{-1}}_{-1.37\times10^{-1}}$  &  88.237  &  $^{+4.80\times10^{-1}}_{-4.80\times10^{-1}}$ \\ \hline 
    \end{tabular}}
    \caption{Best fit orbital parameters from white light curve fitting. We adopt the \firefly results as our system parameters. \eureka and \firefly values are derived from joint fits to both white light curves. Tiberius parameters are derived from a weighted mean of fits to individual light curves.}
    \label{tab:orbital_parameters}
    
\end{table}

\subsection{Final Transmission Spectrum} 
All three independently reduced spectra from above are in agreement, showing no atmospheric features and being statistically consistent with a flat line. The findings reported in the study do not depend upon which reduction pipeline is used. To select the final transmission spectrum for model interpretation, we performed two tests. The first test computed the mean absolute deviation
of each spectrum relative to the averaged spectrum of the three reductions. The purpose of this test was to identify the reduction that is the most representative of the three reductions. The \firefly reduction was favored by this test.  The second test computed the reduced chi-squared relative to a flat line. This test was meant to validate the size of the error bars. The unbinned \firefly transmission spectrum had a reduced chi-squared of 1.015.

\subsection{Planet Validation}

The JWST detection of a transit at the same period, phase, and depth as the TESS TOI eliminates the possibility of a TESS false positive due to a telescope or instrument systematic effect. This leaves only astrophysical sources, such as a background eclipsing binary, as the remaining false positive mechanism. 

\begin{figure}
    \centering
    \includegraphics[width = 0.8\textwidth]{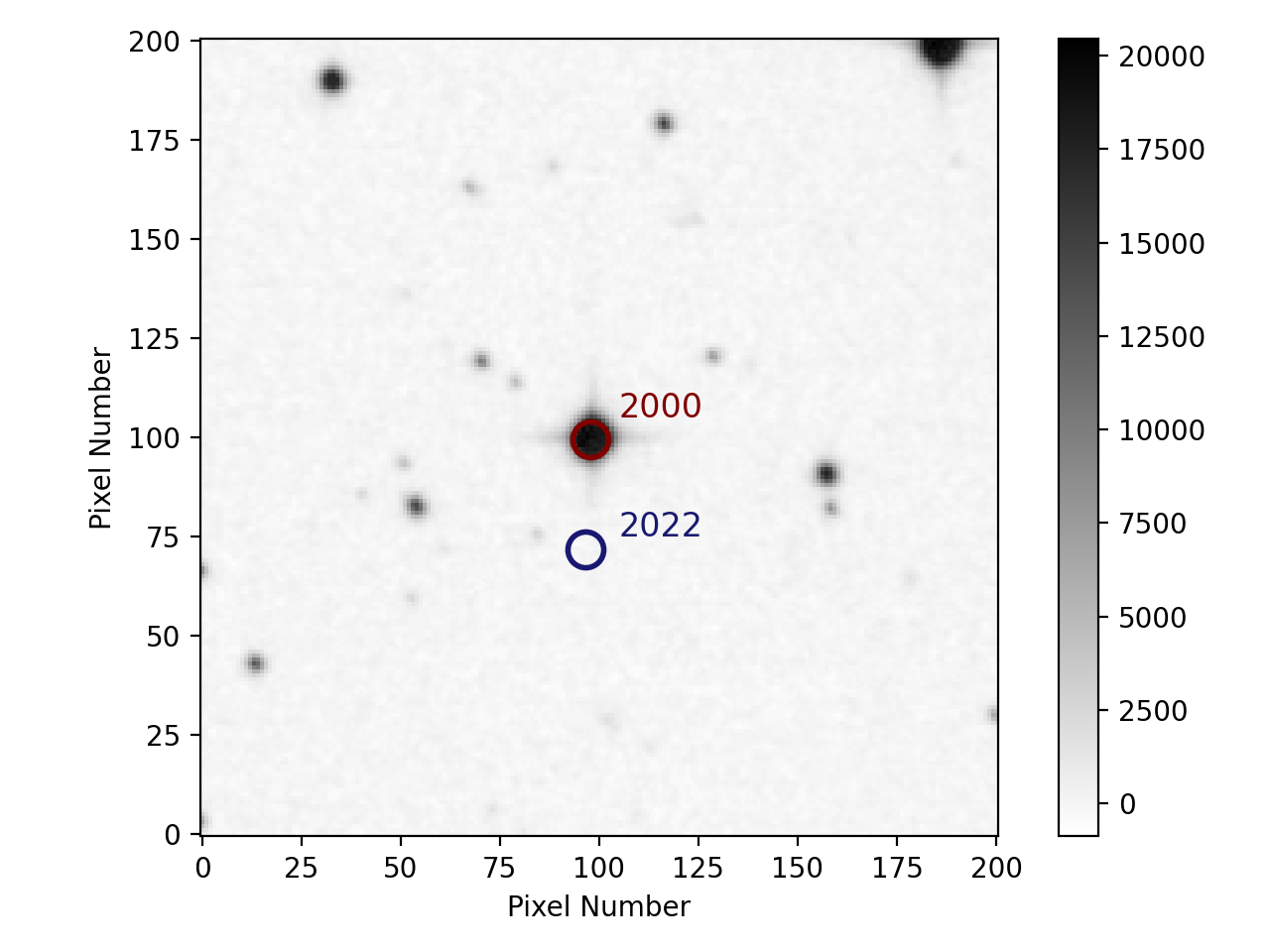}
    \caption{A 3.36" $\times$ 3.36" DSS image centered on LHS~475 taken 1999 June 20.  The red circle depicts the star's J2000 position per Simbad, whereas the blue circle indicates the star's position for the JWST observations in September of 2022.  We see no indication of a background star at the 2022 position that could be the source of the observed transit signal.  For reference, NIRSpec's field of view is 1.6$\times$1.6 pixels on this image.}
    \label{fig:dss}
\end{figure}

Using an archival DSS image of LHS~475, we leverage the star's high proper motion to rule out astrophysical false positives.  The star moves 1.28 arcsec/year \citep[$0.3423$ arcsec/year in RA, $-1.2303$ arcsec/year in Dec;]{Gaia2021}, which corresponds to $\sim$29 pixels in \Cref{fig:dss} from the June 1999 DSS image to our Sep 2022 JWST observation.  The lack of measurable flux at LHS~475's 2022 position enables us to rule out all scenarios involving transits within a potential background system.  Finally, we rule our a stellar binary companion due to the precisely measured Gaia DR3 parallax of 80.1134 mas, which corresponds to a distance of only 12.5 pc.

\subsection{Implications for JWST/NIRSpec Noise Floor}

\begin{figure}
    \centering
    \includegraphics[width = 0.49\textwidth]{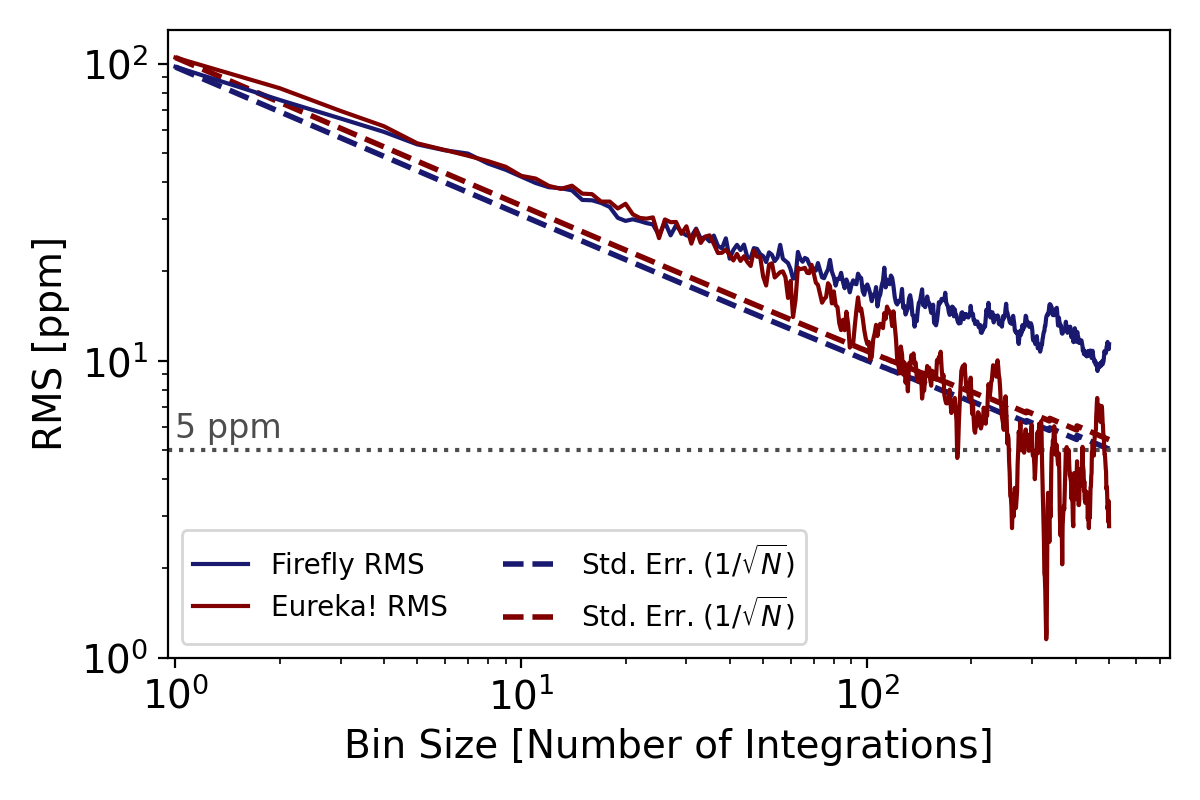}
    \includegraphics[width = 0.49\textwidth]{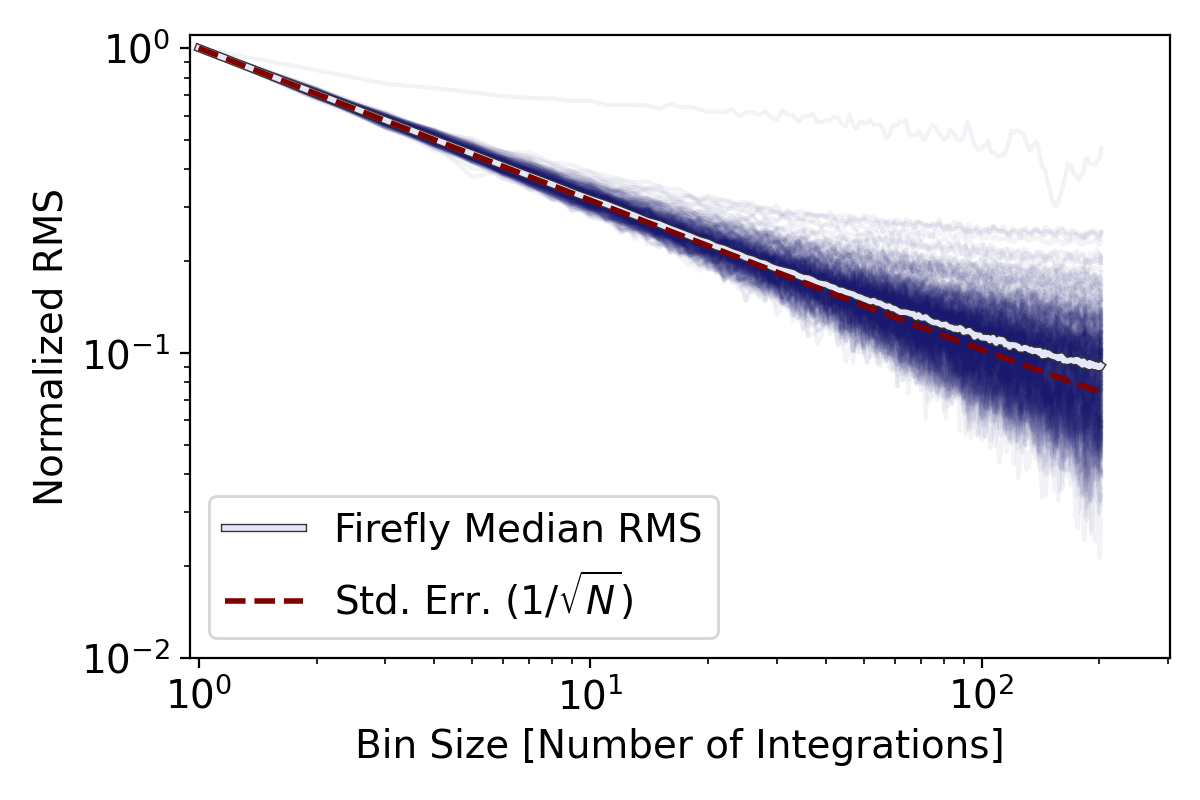}
    \includegraphics[width = 0.49\textwidth]{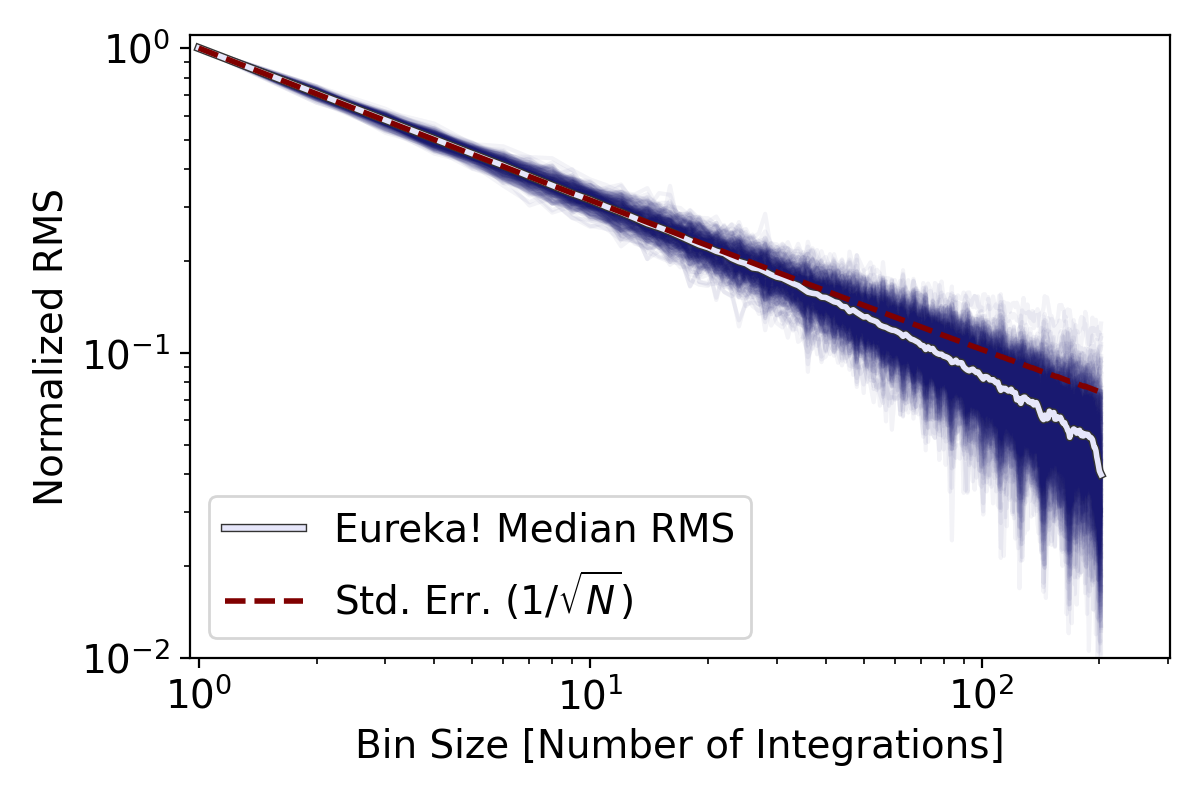}
    \includegraphics[width = 0.49\textwidth]{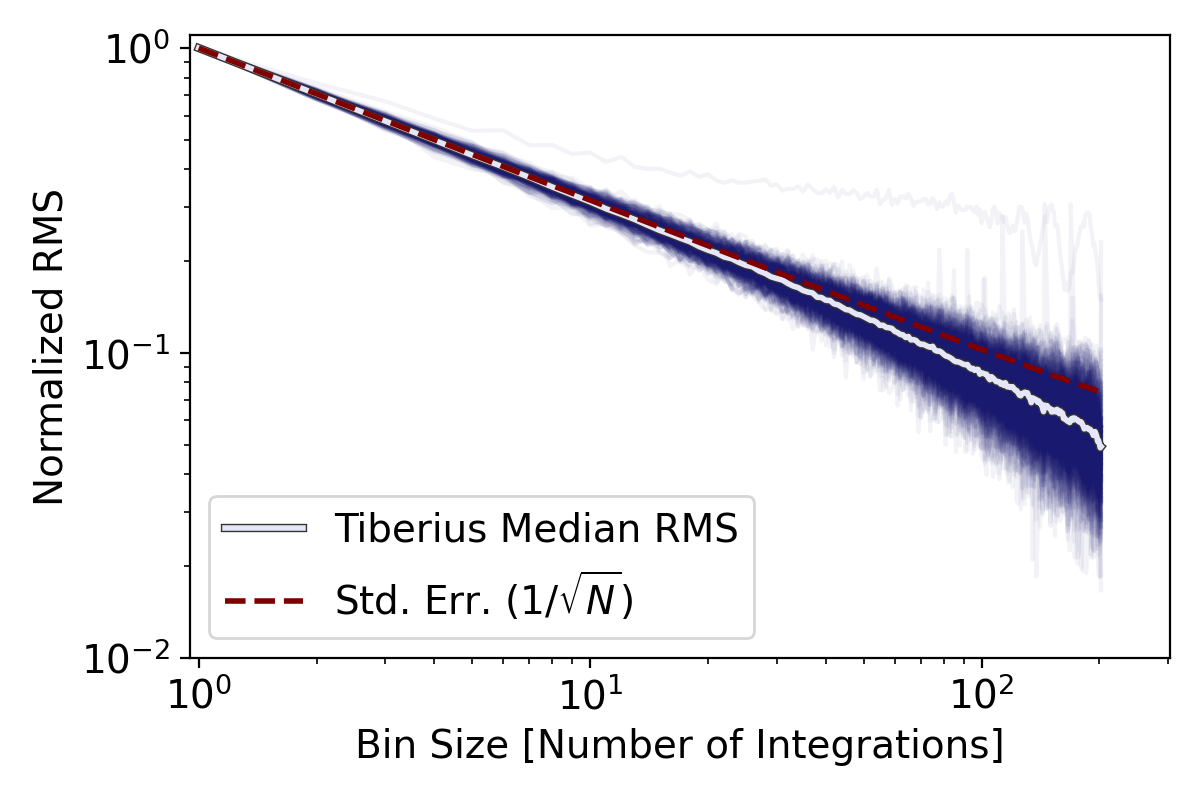}
    \caption{Allan variance plots from the white and spectroscopic light curve fits. Panel (a) illustrates that the white light curve residuals from two of the analyses exhibit some correlated noise at timescales of $<5$ minutes ($<35$ integrations).  This is likely due to uncorrected 1/f noise from the thermal cycling of on-board heaters \citep[][Section 4.5.3]{Rigby2022}.  At longer timescales ($>18$ minutes), the {\eureka} pipeline returns to the expected standard error with RMS values below 10 ppm.  The Tiberius reduction did not sum the flux across both detectors and was not used for this noise floor analysis. 
    The spectroscopic RMS values in panels (b) -- (d) are more consistent with the standard error, thus confirming that the spectroscopic light curves are dominated by white noise.} 
    \label{fig:allan}
\end{figure}

In the interest of exploring the effects of correlated noise and constraining the instrument noise floor, we concatenate residuals from both visits and compute Allan variance plots for the white and spectroscopic light curve fits (see \Cref{fig:allan}).  Using the {\eureka} white light curve data, we find no indication of a noise floor down to 5 ppm; however, we identify correlated noise at timescales of $<5$ minutes.  This timescale is consistent with the thermal cycling of heaters in the ISIM Electronics Compartment, which induces small forces on the telescopes backplane structure \citep{Rigby2022}.  The effect is semi-periodic, the result of several heaters cycling at different frequencies.

\section{Modeling}

With the reduced data and coadded transmission spectrum produced in the previous section, we now use a variety of models to update the state of knowledge on the LHS~475 system. 
We use archival photometry to update the LHS~475 stellar parameters and assess the impact of stellar contamination on the JWST transmission spectrum; we use empirical mass-radius relations to estimate the planet mass given our precise radius measurement; and we fit atmospheric models  to the transmission spectrum to obtain constraints on the possible atmospheric composition of LHS~475b. 

\subsection{Stellar Modeling and Transit Light Source Contamination}

\begin{figure}
    \centering
    \includegraphics[width = \textwidth]{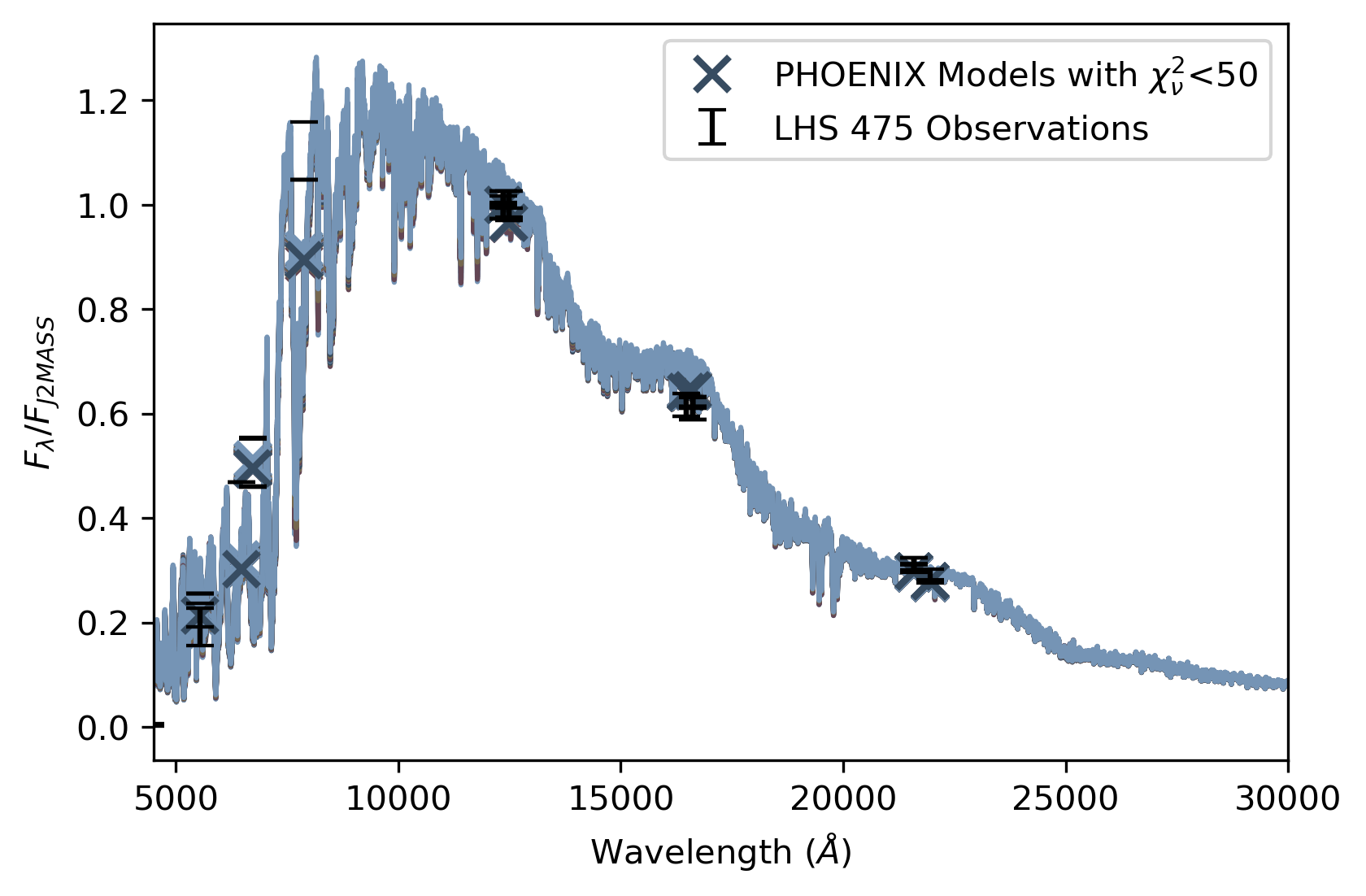}
    \caption{Comparison of our 30 closest matching \texttt{PHOENIX} models ($\chi_\nu^{2}< 50$) to all available archival photometry of LHS 475 from the VizieR Photometry Viewer. These models have $T_{\rm{eff}}$ = 3380 -- 3320 K, log($g$) = 4.7 -- 5.7 g/cm$^{2}$, $M$ = 0.262 $M_{\odot}$.}
    \label{fig:phx}
\end{figure}

We use \texttt{PHOENIX} spectra guided by archival photometry\footnote{For NIRSpec observations, the \texttt{jwst} pipeline requires the flat field step be run for absolute flux calibration. At the time of writing, only ground or dummy frames were available for the three types of NIRSpec flat fields and for the correction applied in the \texttt{photom} step of the \texttt{jwst} Stage 2 pipeline. These ground and dummy frames do not provide high accuracy absolute flux calibration, so we choose to not use our new data for Stellar Modeling at this time.} from the VizieR Photometry Viewer\footnote{\url{http://vizier.cds.unistra.fr/vizier/sed/}} to improve constraints on the stellar parameters. Effective temperature ($T\rm_{eff}$) is a primary driver of spectral shape and so we are able to refine estimates by matching models to the observations at visible and near-IR wavelengths. To do this, we computed a grid of synthetic spectra following similar procedures to those outlined in \cite{Husser2013} with $T\rm_{eff}$ = 3200 -- 3400 K ($\Delta$T = 10 K), log($g$) = 4.5 -- 5.2 dex ($\Delta$log($g$) = 0.1 dex), and M$_{\star}$ = 0.262 M$_\odot$. This parameter space was chosen by expanding around the stellar parameters published in the Tess Input Catalog \citep{Stassun2019}.  For each model, we computed synthetic visible and near-IR photometry over the same wavelengths as the filter profiles for available measurements for LHS~475 and used a reduced $\chi^2$ test to identify the model that most closely matched the observations. The reduced $\chi^2$ test was conducted with both the observations and models normalized to the 2MASS J band flux density value to isolate matching the spectral shape. The fully explored grid yielded $\chi_\nu^2$ values between 6.8 -- 987.2, with 30 models returning similar values less than 50 (Figure \ref{fig:phx}). These models have $T\rm_{eff}$ = 3300 (+80, -30) K, log($g$) = 5.2 $\pm$ 0.5 g/cm$^3$, and M$\star$ = 0.262 M$_\odot$. To determine the radius of the star we scaled all models with $\chi_\nu^2 < 50$ by $R_{\star}^2/dist^2$ until $F_{J2MASS, mod}$ = $F_{J2MASS, obs}$,

\begin{equation}
    R_{\star} = \sqrt{(F_{J2MASS, obs}/F_{J2MASS, mod}) \times dist^2}
\end{equation}

We adopted the Gaia EDR3 distance of 12.481 $\pm$ 0.0065 pc \citep{Gaia2021}, which returned a radius of 0.2789 $\pm$ 0.0014 R$_{\odot}$.

LHS 475 is typical of low-activity M dwarfs in the solar neighborhood. TESS only detected two flares on LHS 475, both with energy below $10^{31}$ erg. The inferred flare rate and other activity diagnostics are all consistent with the general population of relatively inactive M dwarfs in a volume-limited sample \citep{Medina2020}. H$\alpha$ and He I D$_3$ are both in absorption, not emission. Ca II 8542 is relatively deep. 

Fitting models of the Transit Light Source (TLS) effect to the observed and coadded transmission spectrum allows us to assess the degree to which stellar contamination may impact and/or explain any characteristics of the planet's transmission spectrum. The TLS effect can impart slopes and features into the transmission spectrum due to differences in the spot or faculae coverage along the planet's transit chord relative to the average coverage across the visible stellar disk \citep{Rackham2018}. Following the formalism of \citep{Zhang2018}, we calculate the TLS contamination spectrum 
\begin{equation}
\label{eqn:tls}
    \epsilon_{\lambda} = \frac{(1 - f_{\rm spot} - f_{\rm fac}) S_{\lambda,{\rm phot}} + f_{\rm spot} S_{\lambda,{\rm spot}} + f_{\rm fac} S_{\lambda,{\rm fac}}}{(1 - F_{\rm spot} - F_{\rm fac}) S_{\lambda,{\rm phot}} + F_{\rm spot} S_{\lambda,{\rm spot}} + F_{\rm fac} S_{\lambda,{\rm fac}}}
\end{equation}
where $S_{\lambda,{\rm phot}}$, $S_{\lambda,{\rm spot}}$, and $S_{\lambda,{\rm fac}}$ refer to the spectrum of the stellar photosphere, spots, and faculae, respectively, $f_{\rm spot}$ and $f_{\rm fac}$ refer to the spot and faculae projected area covering fractions along the transit chord, and similarly $F_{\rm spot}$ and $F_{\rm fac}$ refer to the spot and faculae projected area covering fractions across the entire visible stellar disk. Thus, $\epsilon_{\lambda}$ is the ratio of the stellar spectrum along the transit chord to the spectrum of the whole disk, and a general model for how the TLS effect contaminates the observed transmission spectrum. Given Equation \ref{eqn:tls} the observed drop in flux that we refer to as the transmission spectrum is simply 
\begin{equation}
    \Delta F_{\lambda,{\rm obs}} = \epsilon_{\lambda}  \left(\frac{R_p}{R_s} \right )^{2}_{\lambda}    
\end{equation}
where the TLS contamination spectrum is multiplied by the wavelength-dependent ``true'' planet transmission spectrum.  
We use the \texttt{Dynesty} nested sampling code \citep{Speagle2020} to infer posterior distributions for the TLS contamination model parameters under the assumption of a wavelength independent planet transmission spectrum. We run the standard nested sampling algorithm \citep{Skilling2004} with 1000 live points until the estimated contribution to the total evidence from the remaining prior volume drops below the threshold of \texttt{dlogz=0.075}. 

\begin{figure}
    \centering
    \includegraphics[width = \textwidth]{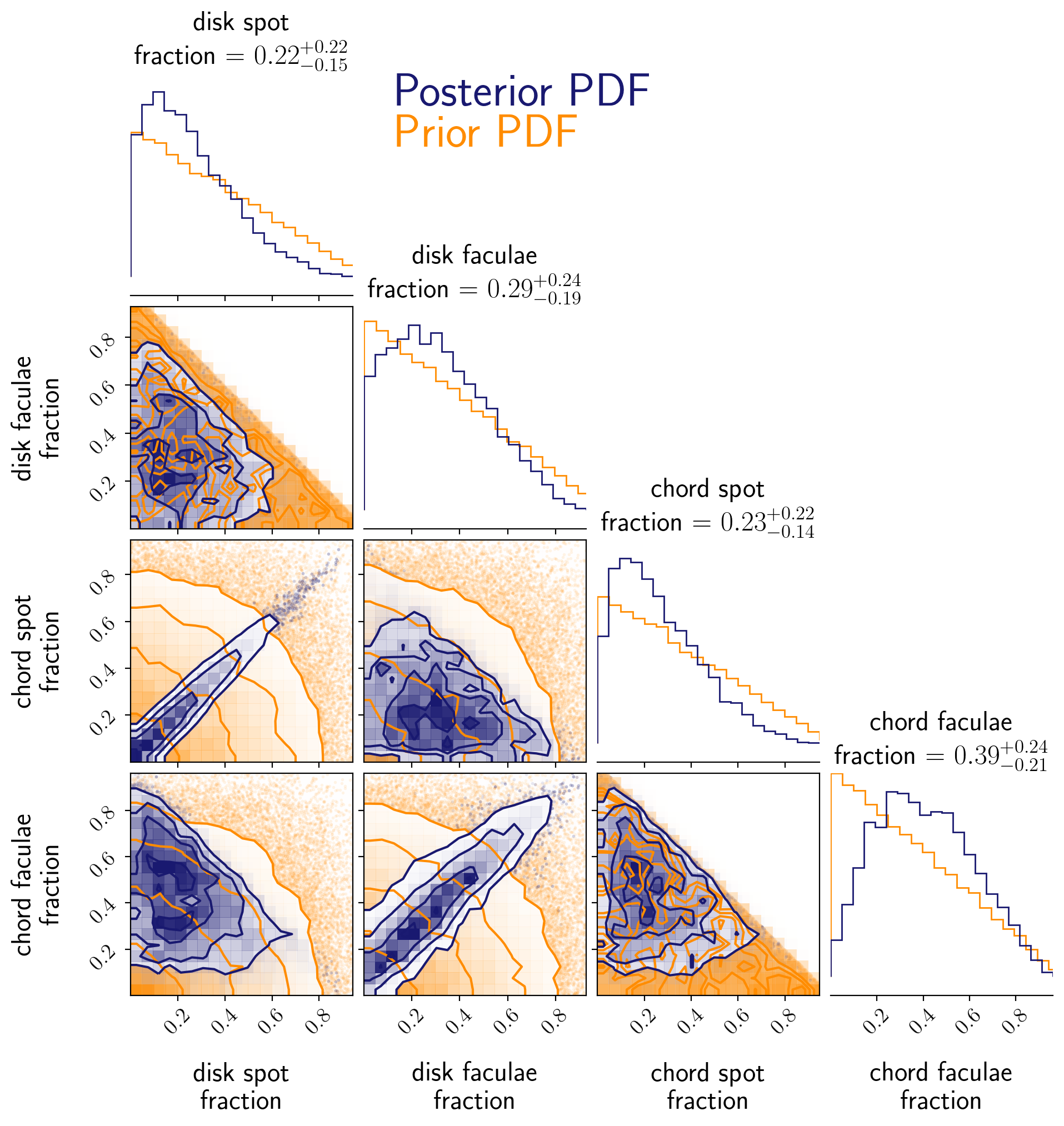}
    \caption{Corner plot comparing the prior (orange) and posterior (dark blue) PDFs for a subset of the fitting parameters in the TLS contamination retrieval. The flat spectrum reveals a consistent spot (and faculae) coverage along the transit chord compared to the full stellar disk.}
    \label{fig:TLS}
\end{figure}

In general, no evidence of TLS contamination is observed in the flat transmission spectrum and the TLS model readily reproduces the featureless spectrum. Figure \ref{fig:TLS} compares the prior and posterior probability distributions for a subset of the TLS model parameters. The inferred posterior distribution for the TLS contamination model generally reproduce the prior distributions, with the exception of the covariance between the spot (faculae) area covering fraction along the transit chord compared to the spot (faculae) covering fraction on the full disk. These two convariance are constrained along a line with a slope of approximately unity, such that the ratio of spot (faculae) covering fraction on the full stellar disk to spot (faculae) covering fraction along the transit chord is $1.005 \pm 0.003$ ($0.948 \pm 0.006$). This implies that---although the exact area covered by spots (and faculae) is not well constrained---at the observed precision there is no evidence of differing spot (or faculae) coverage along the transit chord compared to the average stellar disk. We repeated the same TLS contamination retrieval with the addition of the transit depth measured by TESS in the optical ($978 \pm 73$ ppm) and obtained the same result.   

\subsection{Planet Radius, Mass, and Equilibrium Temperature} 

From the constraint on the white light curve transit depth ($1060 \pm 9$ ppm) and the stellar radius ($\mathrm{R}_s = 0.279 \pm 0.014~\mathrm{R}_{\odot}$), we calculate the planet radius to be $\mathrm{R}_p = 0.991 \pm 0.050~\mathrm{R}_{\oplus}$ ($6319 \pm 318$ km). The $5\%$ radius precision is dominated by uncertainty in the stellar radius. For reference, the preexisting radius constraint from TESS sectors 12-39 was $0.93 \pm 0.70 ~\mathrm{R}_{\oplus}$ \citep{Guerrero2021}.  

Despite the lack of a mass measurement for LHS~475b, we use three different methods to estimate the mass: 1) from the transmission spectrum \citep{deWit2016}, 2) from probabilistic mass-radius-relation \citep{Chen2017}, and 3) from probabilistic bulk density arguments. 
We use atmospheric models over a range of masses compared to the spectroscopic data from NIRSpec/G395H to infer conservative upper and lower limits of the possible mass for LHS~475b. To do so, we employ the forward model framework discussed in the following section.
First, we find the uppermost mass limit by finding the densest planet that could stably support a hydrogen-helium envelope and fit the data. To obtain a reduced-$\chi^2$ $\leq$ 1, we determine that we must consider a mass of 24 $\mathrm{M}_{\oplus}$. Combined with the precise radius constraint, this upper limit mass results in a planetary density of 119 g/cm$^{-3}$, or 6$\times$ that of pure uranium. Given this unrealistic density, we can clearly reject a hydrogen-helium atmosphere around a very dense planet.
On the other hand, to find the lowest mass that is consistent with the NIRSpec data, we instead consider a planet with a very high mean molecular weight atmosphere, but from a reasonably abundant molecule -- that of pure CO$_2$ -- and scale the mass down until we obtain a reduced-$\chi^2$ $\leq$ 1. Under this atmospheric assumption, we find that masses consistent with the data extend down to 0.78 $\mathrm{M}_{\oplus}$. Together this method gives us a range of masses consistent with the observed atmosphere between 0.78 - 24 $\mathrm{M}_{\oplus}$. 

Next,  we use the mass-radius relationship gleaned from the existing population of small M dwarf exoplanets to estimate the planet's mass given our precise radius constraint. Using the \texttt{Forcaster} code's probabilistic mass-radius relationship and mass prediction tool \citep{Chen2017}, we estimate the mass of LHS~475b to be $\mathrm{M}_p = 0.980^{+0.632}_{-0.359} ~\mathrm{M}_{\oplus}$. 

Our third mass estimate leverages recent results on the interior bulk densities among the M dwarf small planet population. Given the radius constraint for LHS~475b, the planet is consistent with the population of M dwarf planets having rocky interior compositions ($1.21 \pm 0.28~\mathrm{R}_{\oplus}$) \citep{Luque2022}. Therefore, if we assume that LHS~475b is indeed a rocky planet with a mean bulk density consistent with the M dwarf rocky planet population ($0.94 \pm 0.13~ \rho_{\oplus}$) \citep{Luque2022}, then we find the planet mass to be $\mathrm{M}_p = 0.914 \pm 0.187 ~\mathrm{M}_{\oplus}$. If we consider that instead the planet were in the population of lower density water worlds (with 50\% water, 50\% rock interiors), then this would ultimately have ramifications for the scale height and water content of the atmosphere \citep[e.g.,][]{Turbet2019inflation}, that are inconsistent with the featureless transmission spectrum that we measured. Therefore, it is likely that \planetname has a mass that is consistent with a rocky mean bulk density. 
In the atmospheric models that follow, we assume the planet is consistent with the population with rocky interiors and use the corresponding mass $\mathrm{M}_p = 0.914 \pm 0.187 ~\mathrm{M}_{\oplus}$, which is consistent with our previous estimates, albeit with a tighter constraint. 

We update \planetname's zero bond albedo equilibrium temperature to $586 \pm 12$ K (assuming uniform heat redistribution). In the limit of instant re-radiation expected from a planet with a tenuous or nonexistent atmosphere, the estimated day side brightness temperature is $748 \pm 16$ K. These updates may aid in the planning of any future secondary eclipse observations of \planetname. 

\subsection{Atmospheric Modeling}

We use atmospheric radiative transfer models to simulate the transmission spectrum of \planetname for comparison with our JWST observations. In the next section, forward models of single-composition end-member atmospheres and archetypal atmospheres are used to illustrate the atmospheric compositions that are consistent with our observed data. Then, retrieval models are used to simulate a broad range of atmospheric compositions to place constraints on key atmospheric parameters given the precise, yet featureless transmission spectrum. 

\subsubsection{Forward Modeling}

We use the forward modeling capabilities of two different open-source atmospheric radiative transfer codes, \texttt{PICASO} \citep{Batalha2019} and \texttt{CHIMERA} \citep{Line2013b,Line2014-C/O}, to explore the plausibility of various atmospheric archetypes. We compute each model atmosphere for a planet mass consistent with a rocky mean bulk density, $\mathrm{M}_p = 0.914 ~\mathrm{M}_{\oplus}$, a planetary radius of $\mathrm{R}_p = 0.991 ~\mathrm{R}_{\oplus}$, and a stellar radius $\mathrm{R}_s = 0.279 ~\mathrm{R}_{\odot}$, and a planetary equilibrium temperature of $\mathrm{T}_{eq}$ = 600 K, as above.
In each case, we compare the modeled transmission spectrum to the NIRSpec/G395H data for LHS~475b from 2.9 -- 5.3 {\micron} and compute the reduced-$\chi^2$ between the modeled spectrum and the data. 

For the \texttt{CHIMERA} models, we compute chemically consistent atmospheric mixing ratios for 1$\times$, 10$\times$, 100$\times$, and 1000$\times$ solar metalicities, with a solar C/O ratio. \texttt{CHIMERA} uses a preset grid of atmospheric molecular abundances along temperature-pressure profiles, metallicity, and C/O ratio generated from the NASA CEA code \citep{McBride1996}. For the temperature-pressure profile, the code uses the five-parameter, double gray, one-dimensional parametrization of \citep{Guillot2010}, where we input a planetary equilibrium temperature of 600 K. For these \texttt{CHIMERA} models, we include opacity from H$_2$O, CH$_4$, CO, CO$_2$, NH$_3$, N$_2$, HCN, H$_2$S, H$_2$/He CIA \citep{Freedman08,Freedman2014}, and Rayleigh scattering from H$_2$. We consider simplistic cloudy hydrogen-dominated atmospheric models with \texttt{CHIMERA} by computing a cloud-top pressure for a grey absorbing cloud. We generate atmospheric transmission models with the correlated-k method of radiative transfer and bin the resulting model to the data before calculating our reduced-$\chi^2$.

\begin{figure}[t]
    \centering
    \includegraphics[width=0.99\linewidth]{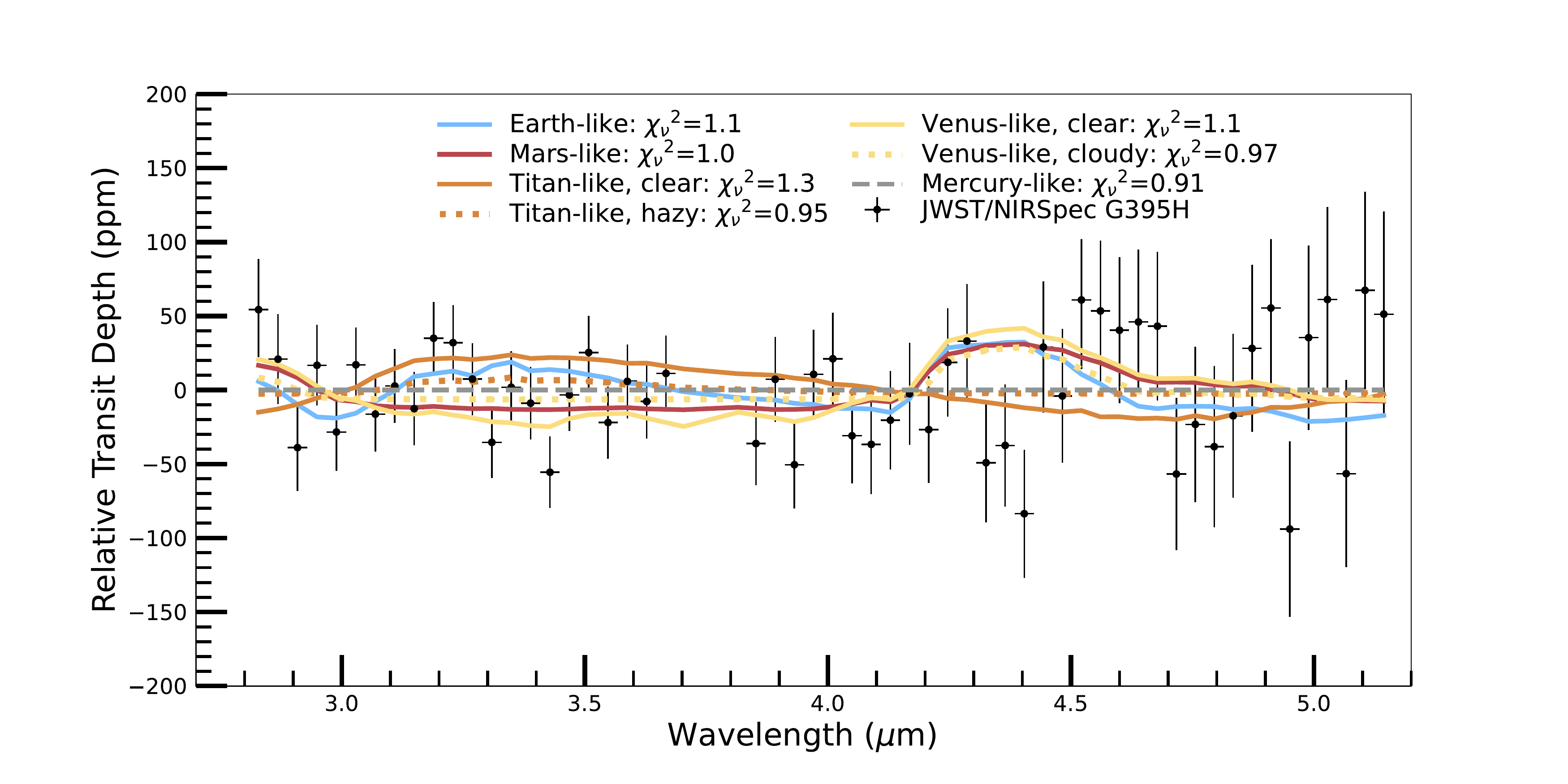}
    \caption{\label{fig:solar_system_models}{
    Final, binned spectrum (black points) compared to atmospheric models with compositions of the Solar System terrestrial planets (coloured lines). Our data, to weakly rule out Earth composition (blue solid), clear Titan composition (orange solid), and clear Venus composition atmospheres (yellow solid). However, the data are all consistent within error to that of a hazy Titan composition with a haze-top at 0.01 mbar (dotted orange), a cloudy Venus composition with a cloud-top at 1 mbar (dotted yellow), and a Mars composition atmosphere (red solid), as well as that of an airless body, like Mercury (grey dotted line).
    }}
\end{figure}

For the \texttt{PICASO} models, we generate simplified end-member atmospheric compositions with isothermal temperature-pressure profiles. We set a pressure grid which ranges from 1 $\mu$bar to 100 bar, and then set an isothermic temperature at the equilibrium temperature of 600 K. For the models shown in Figure \ref{fig:spectrum_zoom}, each atmosphere consists solely of either H$_2$O, CO$_2$, CH$_4$, or as in the case of the Earth-like atmosphere, follows the atmospheric abundances of Earth above the water cold-trap, with 78\% N$_2$, 21\% O$_2$, 0.9\% Ar, 416 ppm CO$_2$, 524 ppm He, and 187 ppm CH$_4$. For the models shown in Figure \ref{fig:solar_system_models}, we generate individual pressure grids with an upper bound according to the terrestrial body's surface pressure (e.g., the Earth-like model has an upper atmospheric pressure bound of 1 bar; the Venus-like model has an upper pressure bound of 90 bar). We assume isothermal temperature profiles (at 600 K) with atmospheric abundances fixed to the composition of each Solar System body above any cold trap. For the cloudy Venus and hazy Titan cases, we implement a simple grey absorbing cloud at the pressure level according to the Venus cloud-top (1 mbar) and the Titan haze-top (0.01 mbar). The opacity database is resampled to R=10,000 and is taken from \citep{Batalha2020}. Models are then binned to the data for reduced-$\chi^2$ comparison. 

We strongly rule out clear atmospheres of 1$\times$ to 100$\times$ solar, with reduced-$\chi^2$s $\geq$ 9, or over 10$\sigma$. Given the mass estimate analysis above, even with the uncertain planetary mass, we are able to reject low ($\le$100$\times$ solar) atmospheres. To obtain a reduced-$\chi^2$ $\sim$ 1 in cloudy low-metallicity atmospheres, we must insert an opaque cloud deck with cloud-top pressure between 0.5 and 1 $\mu$bar, which can be discarded as unrealistic given the lack of cloud-forming material at such low pressures. Each of these cases represents a hydrogen-rich atmosphere around a rocky, 600 K planet, which would not be stable against escape over the lifetime of the system, and thus our ability to reject them is not unexpected. 

For the 1000$\times$ atmosphere, we calculate a reduced-$\chi^2$ to the data of 1.5, which weakly rules out this scenario to 2.5$\sigma$. The pure methane atmosphere is rejected with a reduced-$\chi^2$=2.3, or 5$\sigma$. Both the end-member atmospheric compositions of a pure steam or Earth-like atmospheric abundances are weakly disfavored at $\gtrsim$1$\sigma$. A pure 1 bar carbon dioxide atmosphere or no atmosphere at all are preferred but not statistically distinguishable from each other. For the Solar System terrestrial archetype atmospheres shown in Figure \ref{fig:solar_system_models}, we also weakly disfavor ($\gtrsim$1$\sigma$)  clear Venus, Titan, or Earth-like atmospheres, but cannot statistically distinguish between a thin Mars-like atmosphere, a hazy Titan-like atmosphere, or a cloudy Venus, as consistent with the retrieval modeling shown in Figure \ref{fig:retrieval_pair} and discussed below.

\subsubsection{Retrieval Modeling} 

We use two different atmospheric retrieval codes---\texttt{smarter} and \POSEIDON---to explore the range of atmospheric properties that are consistent with, or ruled out, by LHS~475b's transmission spectrum. 

\paragraph{Retrievals with \texttt{smarter}}

The \smarter retrieval code \citep{Lustig-Yaeger2020thesis, Lustig-Yaeger2022} couples line-by-line radiative transfer calculations from the Spectral Mapping Atmospheric Radiative Transfer forward model (\texttt{smart} \citep{Meadows1996}) to the \texttt{dynesty} nested sampling Bayesian inference code \citep{Speagle2020} to retrieve planetary and atmospheric parameters that are consistent with the JWST observations. We assume an isothermal temperature-pressure profile and evenly-mixed gas volume mixing ratios. We calculate line absorption coefficients for gaseous molecules using the \texttt{lblabc} code \citep{Meadows1996} with inputs from the HITRAN2016 line list  \citep{Gordon2017HITRAN2016}. To speed up the retrieval calculations, absorption coefficients are produced for an isothermal temperature of 550 K and resampled to a fixed wavenumber resolution of 0.25 cm$^{-1}$. Our tests that relaxed these assumptions on the line absorption coefficients resulted in negligible errors relative to the measurement uncertainties. 

Our nominal \smarter retrieval setup uses 9 free parameters that include the $\log_{10}$volume mixing ratios for the molecules \ce{H2O}, \ce{CH4}, \ce{CO2}, and \ce{CO}, along with the reference radius of the planet ($R_{p, \rm{ref}}$) at the spectral continuum (which is interpreted as either a cloud-top or the solid-surface), the atmospheric pressure at the reference radius ($P_0$), the isothermal temperature ($T_0$), the planet mass ($M_p$), and the mean molecular weight (MMW) of the bulk atmospheric composition ($\mu$). We impose uninformative flat priors on the gases within the interval $\mathcal{U}(-12,0)$ $\log_{10}(\rm VMR)$, the radius within $\pm 10 \%$ of the white light radius constraint, the apparent surface pressure $P_0 \sim  \mathcal{U}(-6, 1)$ $\log_{10}$(bar), and the isothermal temperature $T_0 \sim \mathcal{U}(200, 900)$ K. The total atmospheric MMW is calculated self-consistently from the gases included in the retrieval plus an unknown, agnostic background gas that fills the remaining volume of the atmosphere after the other gases are accounted for. The agnostic background gas has a molecular weight sampled from a flat prior distribution $\mu \sim \mathcal{U}(2.5, 50.0)$ g/mol. This covers a range in MMW from a low mass solar composition mixture of \ce{H2}+He to high mass, simple molecules such as \ce{CO2} and \ce{O3}. While this model construction is similar to other retrievals that assume a known background gas such as \ce{H2}+\ce{He} or \ce{N2}, using a flat prior on the molecular weight of the background gas eliminates a strong implicit prior on the total atmospheric MMW (which is strongly biased to that of the assumed background gas). We assume that the planet possesses a rocky interior composition, as previously discussed, and sample planet masses from a normal distribution  $\mathrm{M}_p \sim \mathcal{N}(0.914, 0.187) ~\mathrm{M}_{\oplus}$. 

We run \smarter retrievals using the \texttt{dynesty} code with the standard nested sampling algorithm \citep{Skilling2004} and fit the final coadded transmission spectrum from the \firefly reduction binned to a fixed resolution of $\Delta \lambda = 10$ nm. We use 600 live points and run the model until the estimated contribution to the total evidence from the remaining prior volume drops below the threshold of \texttt{dlogz=0.075}. To obtain additional posterior samples that effectively reduces the numerical sampling errors in the final visualization of the posteriors, we run an MCMC chain using \texttt{emcee} \citep{emcee2013}. The MCMC is run with 135 walkers for 1000 steps and is initialized using points from the equally weighted \texttt{dynesty} posterior. The resulting MCMC chain requires no iterations to be removed for the burn-in and the \texttt{emcee} posteriors agree with the \texttt{dynesty} posteriors to within the finite sampling uncertainty. Our final posteriors are constructed by combining the list of samples obtained with the two inference codes. 

\begin{figure}
    \centering
    \includegraphics[width = \textwidth]{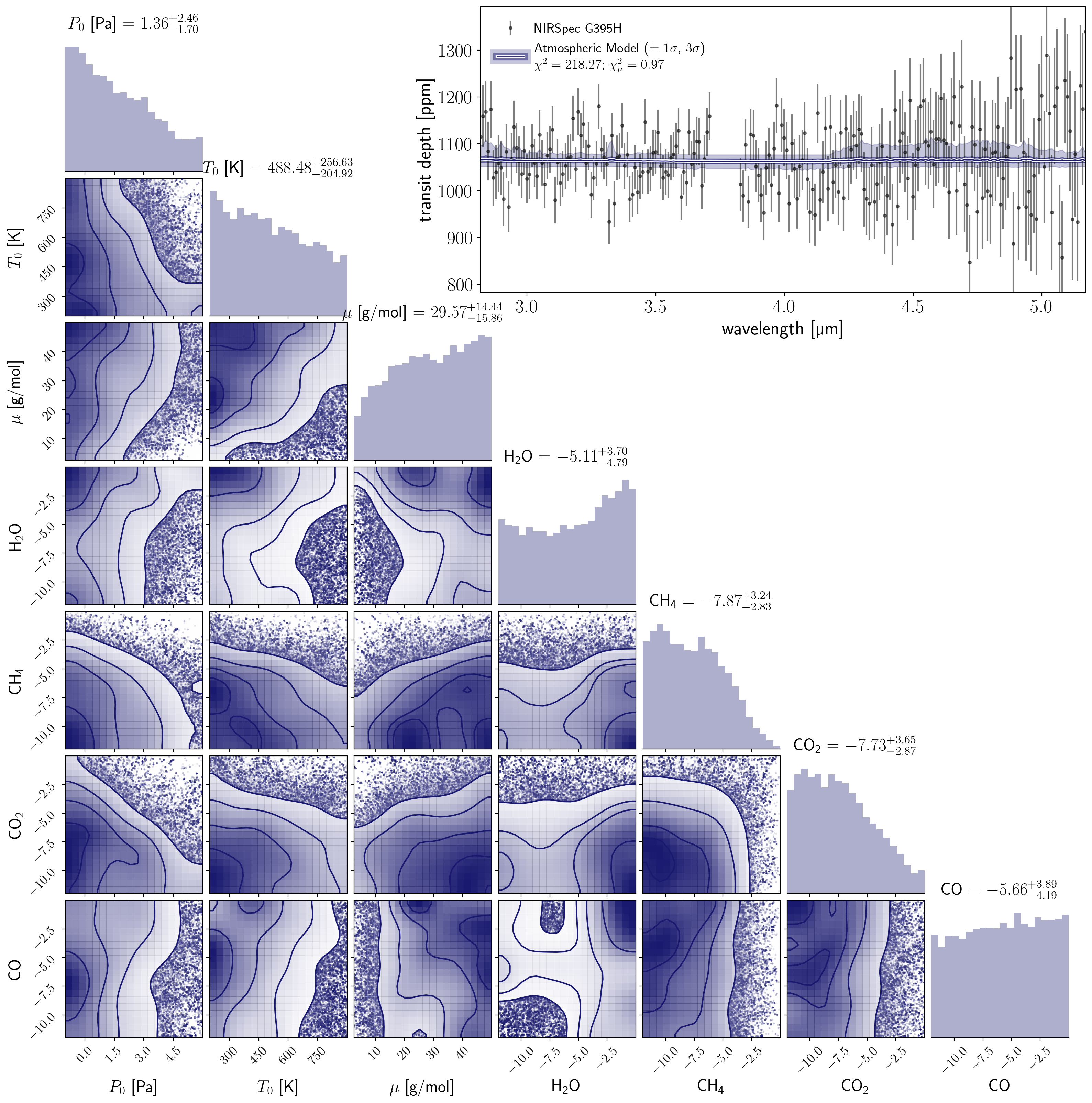}
    \caption{Corner plot showing the 1D and 2D marginalized posterior probability distribution for a subset of the \texttt{smarter} model parameters. The upper right axis shows the 1$\sigma$ and 3$\sigma$ envelope around the median retrieved spectrum, which corresponds to the multidimensional posterior PDF projected onto the observed spectrum.  Disfavored atmospheres are thick (large $P_0$), hot (large $T_0$), and composed primarily of light molecules (low $\mu$).}
    \label{fig:retrieval_corner}
\end{figure}

Figure \ref{fig:retrieval_corner} shows the posterior PDFs from our nominal \texttt{smarter} retrieval along with an overview of spectral models sampled from the posterior. Although a large swath of terrestrial atmospheric parameter space remains allowed given the observations, a non-negligible subset of models are disfavored and provide us with insights into the nature of the planet. 
Figure \ref{fig:retrieval_pair} highlights these constraints and includes the isothermal scale height for atmospheres contained in the posterior distribution. Scale height calculations are performed in a post-processing step after running the retrieval to compress the degeneracies between planet gravity (fit in terms of planet radius and mass), isothermal temperature, and mean molecular weight into a single representative value for the atmosphere's vertical extensiveness. 
In general, atmospheric characteristics that tend towards increasing the scale height of the atmosphere are disfavored, including high temperatures and low mean molecular weight bulk atmospheric compositions, particularly for atmospheres with apparent surface pressures ${\gtrsim} 10$ mbar (1000 Pa). Conversely, compact atmospheres with small scale heights---due to high mean molecular weight molecules or cool temperatures---are allowed across the full range of apparent surface pressures explored. These two general characteristics yield a preference for extremely low apparent surface pressures of ${\sim}1$ $\mu$bar. 
High abundances of \ce{CO2} and \ce{CH4} can be ruled out in the thick and extended atmospheric scenarios. While \ce{CH4} can be ruled out in a relatively low MMW \ce{CH4}-dominated atmosphere (16 g/mol), \ce{CO2} is more difficult to rule out in a heavier \ce{CO2}-dominated atmosphere (44 g/mol). 

The \ce{H2O} marginalized posterior shows a slight uptick towards large VMRs due to a small rise in the spectrum at the blue end (${<}3$ \micron{}) where there is a \ce{H2O} band. However, we caution that since a flat line model fit provides a $\chi^2 \approx 1$, the retrieval is inherently overfitting and cannot lead to a statistically significant detection of molecular absorption from these data. To emphasize this point we fit the spectrum with a generalized Gaussian model (plus a flat transit depth component) \citep[e.g.][]{ERSFirstLook} as a minimally parametric stand-in model for any molecular absorbers not included in the retrieval. This model recognizes the same blue end slope in its maximum likelihood solution, but is disfavored relative to the best fitting flat line at $3.1 \sigma$, further indicating that the ``feature'' is consistent with noise. 

We also run a series of \texttt{smarter} retrieval models with the same setup as previously described except with single gas compositions. From the posterior distributions, we derive the maximum size of molecular absorption features such that any larger and they would have been detected in the spectrum. At 3$\sigma$ confidence, we rule out \ce{H2O} absorption features larger than 61 ppm at 2.8 \micron{}, \ce{CH4} features larger than 38 ppm at 3.3 \micron{}, \ce{CO2} features larger than 49 ppm at 4.3 \micron{}, and \ce{CO} features larger than 62 ppm at 4.6 \micron{}. 

\paragraph{Retrievals with \POSEIDON}

\POSEIDON \citep{MacDonald2017} is an atmospheric retrieval code that has been widely applied to interpret transmission spectra of giant exoplanets. \POSEIDON also supports retrievals of terrestrial exoplanets \citep{Kaltenegger2020,Lin2021}, which we here apply to LHS~475b's transmission spectrum. The most up-to-date description of \POSEIDON's radiative transfer technique, forward atmospheric model, and opacity sources is contained in \citep{MacDonald2022}. We explore the range of possible atmospheres for LHS~475b using the nested sampling algorithm \texttt{PyMultiNest} \citep{Feroz2009,Buchner2014}.

We employ a 9-parameter \POSEIDON retrieval configuration. We compute transmission spectra at a spectral resolution of $R =$ 20,000 from 2.6--5.3\,\micron (using cross sections resampled from a high-resolution wavenumber grid with 0.01 cm$^{-1}$ spacing). Our model atmospheres cover $10^{-7}$--$10$\,bar with 100 layers spaced uniformly in log-pressure. We assume 1D plane-parallel atmospheres with an isothermal pressure-temperature profile, uniform-in-altitude gas volume mixing ratios, and that hydrostatic equilibrium and the ideal gas law hold throughout the atmosphere. The stellar radius is fixed to $R_{\rm{s}} = 0.279 ~\mathrm{R}_{\odot}$. The atmospheric structure and composition are thus described by 7 quantities: the isothermal temperature, $T$, the atmospheric radius at the 1\,bar apparent surface pressure level, $R_{\rm{p, \, ref}}$, and the volume mixing ratios of \ce{H2}, \ce{H2O}, \ce{CH4}, \ce{CO2}, and \ce{CO}. We prescribe \ce{N2} as a spectrally inactive filler gas, which allows the mean molecular weight to vary in a similar manner to the \smarter retrievals, except bounded within the simplex of gas weights included in the model. We also fit for the pressure of an opaque surface (or cloud), $P_{\rm{surf}}$, and the gravitational field strength at the pressure level corresponding to the observed planet radius ($r = 0.991 ~\mathrm{R}_{\oplus}$), $g$.  Our priors for the non-mixing ratio parameters are as follows: $T \sim \mathcal{U}$ (200\,K, 900\,K), $R_{\rm{p, \, ref}} \sim \mathcal{U}$ (0.9\,$R_{\rm{p}}$, 1.1\,$R_{\rm{p}}$), $\log_{10} P_{\rm{surf}} \sim \mathcal{U}$ (-7, 1) (units of bar), and $\log_{10} g \sim \mathcal{N}$ (2.960, 0.099$^2$) (units of cm s$^{-2}$). The Gaussian prior on $\log_{10} g$ arises from error propagation from the uncertainties on $R_{\rm{p}}$ and $M_{\rm{p}}$ --- with the latter uncertainty assuming the same rocky interior assumption as the other models. We use 4,000 \texttt{PyMultiNest} live points during each retrieval.

\begin{figure}
    \centering
    \includegraphics[width = \textwidth]{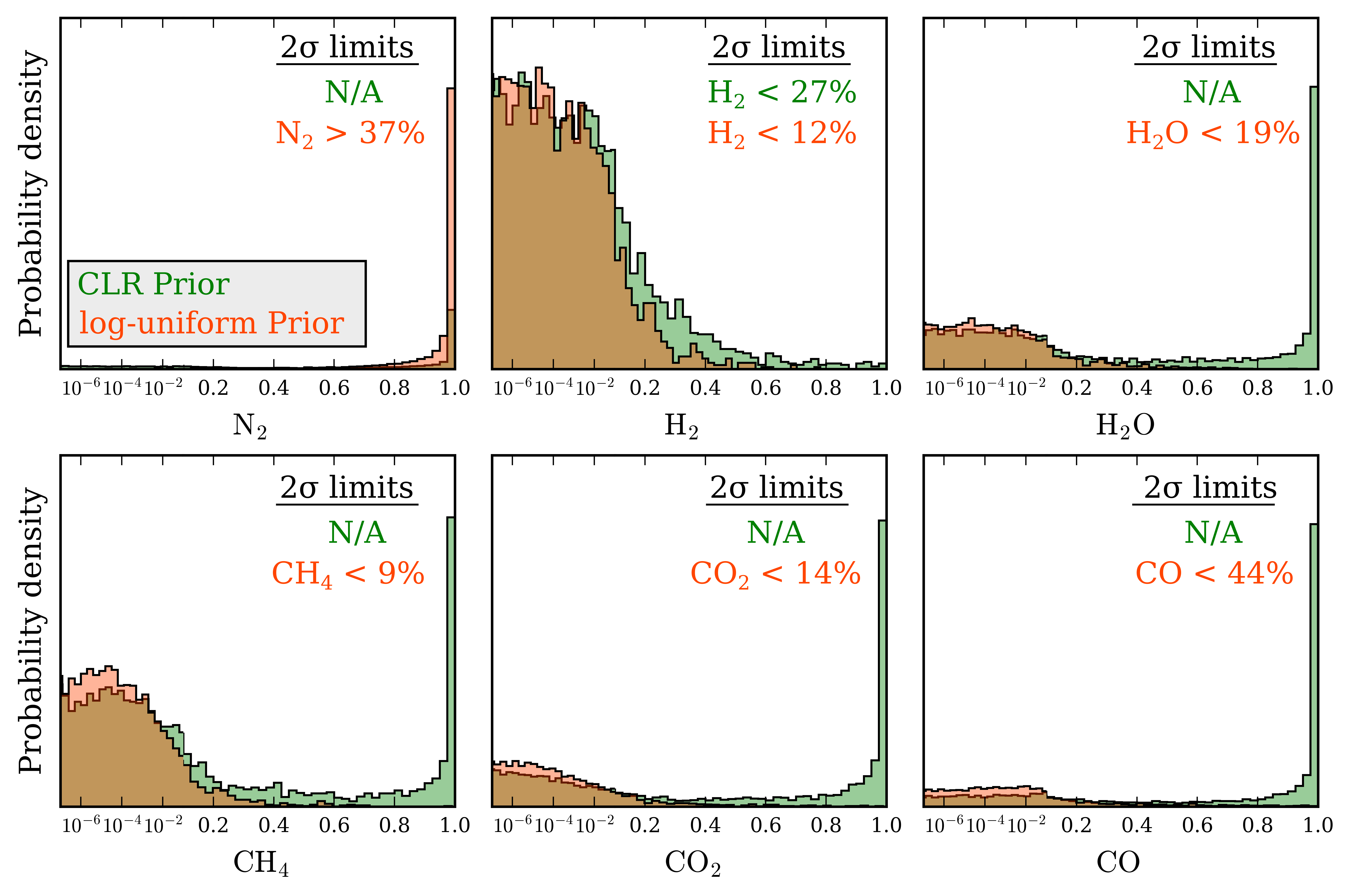}
    \caption{Retrieved volume mixing ratios from the \POSEIDON retrievals of LHS~475b's transmission spectrum. Two retrievals with different prior treatments for the atmospheric composition are overplotted: centered log-ratio (CLR) transformed abundances with \emph{a priori} unknown composition (green); and log-uniform abundances assuming an N$_2$-dominated atmosphere (orange). Statistical 2$\sigma$ upper and lower limits are annotated (or `N/A' if unconstrained). Both retrievals rule out H$_2$-dominated atmospheres. The log-uniform retrieval finds upper limits on H$_2$O, CH$_4$, CO$_2$, and CO due to the assumption that N$_2$ dominates the atmosphere, while the agnostic CLR treatment does not find upper limits for their abundances. For clarity in viewing upper limits, we switch from a logarithmic to linear x-axis at a mixing ratio of 10\%. The probability densities for the linear histogram bins are renormalized to match the probability density of the nearest logarithmic bin left of the 10\% boundary.}
    \label{fig:CLR_vs_log}
\end{figure}

We explore two distinct prior treatments for the atmospheric gas mixing ratios during our \POSEIDON retrievals. Our first approach parameterizes the mixing ratios of \ce{H2}, \ce{H2O}, \ce{CH4}, \ce{CO2}, and \ce{CO} with priors uniform-in-the-logarithm, $\log_{10} X_i \sim \mathcal{U}$ (-12, 0), with the remainder of the atmosphere filled with \ce{N2} ($X_{\rm{N_2}} = 1 - \sum_i X_i$). Any samples requiring negative \ce{N2} mixing ratios are rejected. This `log-uniform' mixing ratio prior is the standard method used for giant exoplanet retrievals, albeit with \ce{H2} + \ce{He} assumed as the filler gas. However, this approach implicitly places a strong prior favouring high abundances for the filler gas \citep{Benneke2012}. For small planets such as LHS~475b, where we do not know \emph{a priori} which gas dominates the atmosphere, one may prefer an agnostic prior that treats all $n$ gases equally. Instead of the agnostic background gas employed by \smarter to resolve this assumption, our second approach with \POSEIDON uses the centred log-ratio transformation (CLR) \citep{Aitchison1986} of the mixing ratios as free parameters: $\xi_i = \ln (X_i / g(\boldsymbol{X}))$, where $g(\boldsymbol{X})=\left(\prod_{j=0}^{n} X_j\right)^{1/n}$. For $n = 6$ gases with a minimum mixing ratio of $X_{\rm{min}} = 10^{-12}$, we ascribe a uniform prior on the 5 CLR variables: $\xi_i \sim \mathcal{U}$ (-20.996, 22.105) --- for our \POSEIDON model, these correspond to \ce{H2}, \ce{H2O}, \ce{CH4}, \ce{CO2}, and \ce{CO}. The upper limit corresponds to the $i^{\rm{th}}$ gas ($i = 1...5$) dominating the atmosphere and all other gases having $X_{j \neq i} = X_{\rm{min}}$, while the lower limit corresponds to $X_{i} = X_{\rm{min}}$ and the other gases equally filling the remainder of the atmosphere. Since $\sum_{i=0}^n \xi_i = 0$ (which automatically ensures $\sum_{i=0}^n X_i = 1$), we use a numerical rejection scheme to ensure that $\xi_0$ (corresponding here to \ce{N2}) falls within the allowed prior range for the other $\xi_i$. The results for the CLR approach are permutation invariant, so switching which gas corresponds to $i = 0$ does not alter the results.

\begin{figure}
    \centering
    \includegraphics[width = \textwidth]{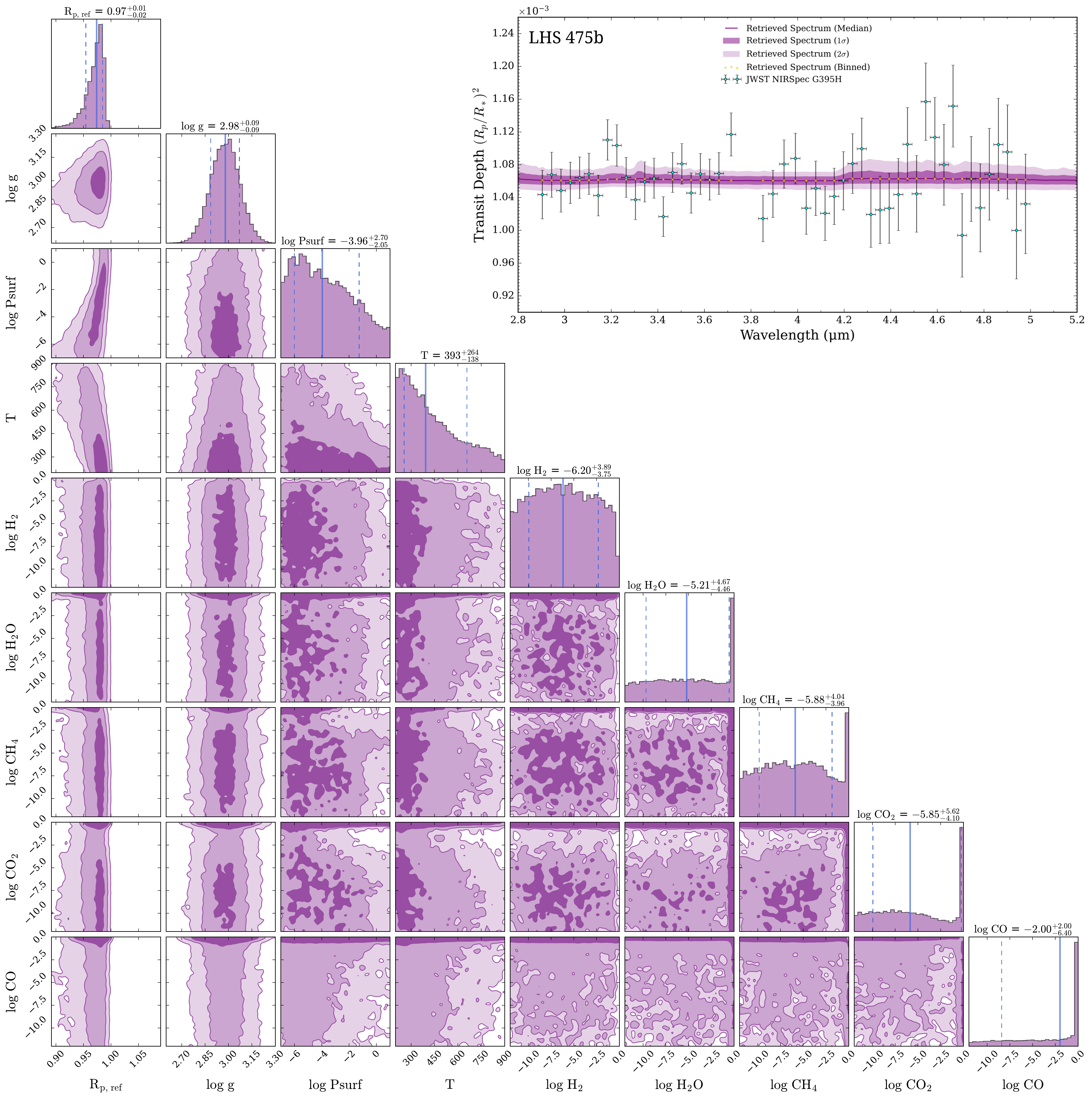}
    \caption{Corner plot showing the 1D and 2D marginalized posterior probability distributions from the \POSEIDON retrieval using CLR mixing ratio parameters. The units are: $R_{\rm{p, \, ref}}$ ($\mathrm{R}_{\oplus}$), $g$ (cm s$^{-2}$), $P_{\rm{surf}}$ (bar), and $T$ (K). The inset shows the corresponding retrieved transmission spectrum model (1$\sigma$ and 2$\sigma$ confidence regions) compared to the NIRSpec G395H observations. The solution rules out H$_2$-dominated atmospheres (to $>$ 5$\sigma$) and thick atmospheres ($P_{\rm{surf}} \gtrsim$ 10\,mbar) dominated by CH$_4$ (to 3$\sigma$).}
    \label{fig:retrieval_corner_2}
\end{figure}

We find that the derived constraints on LHS~475b's atmosphere are sensitive to the choice of mixing ratio prior. Figure~\ref{fig:CLR_vs_log} compares the retrieved abundances from \POSEIDON for the CLR and log-uniform mixing ratio priors. Both approaches rule out H$_2$ dominated atmospheres: log (H$_2$) $<$ 27\% for CLR priors vs. log (H$_2$) $<$ 12\% for log-uniform priors (both 2$\sigma$ upper limits). However, the log-uniform retrieval also infers upper limits on the abundances of H$_2$O, CH$_4$, CO$_2$, and CO --- ranging from $<$ 9\% to $<$ 44\% --- which arise from the built in prior bias towards N$_2$ being the background gas. The CLR prior, in contrast, recognizes that these heavier gases all provide reasonable explanations for LHS~475b's flat transmission spectrum due to their high mean molecular weight --- consistent with the \smarter retrieval which also ruled out low mean molecular weights. However, even with the CLR prior, certain atmospheric scenarios are still disfavored. By examining Figure~\ref{fig:retrieval_corner_2}, which shows the full \POSEIDON posterior for the CLR prior, one can see that CH$_4$ dominated atmospheres with surface pressures $\gtrsim$ 10\,mbar are ruled out to 3$\sigma$ confidence. In other words, our retrieval accounting for the \emph{a priori} unknown background gas confirms the result from our forward modelling analysis that thick, pure CH$_4$ atmospheres with $P_{\rm{surf}} \geq$ 1\,bar are strongly ruled by our LHS~475b transmission spectrum.

\bmhead{Acknowledgments}

\paragraph{Data Availability:} The data used in this paper are from the JWST Cycle 1 General Observer program 1981 and are publicly available on the Mikulski Archive for Space Telescopes (https://mast.stsci.edu). Fully reduced data products from this paper will posted on the Zenodo long term public archive upon acceptance. 

\paragraph{Code Availability:} The codes used throughout this work for data analysis, atmospheric modeling, and manuscript preparation are as follows: 
Astropy \citep{astropy,astropy2}, Batman \citep{batman2015}, \texttt{CHIMERA} \citep{Line2013b,Line2014-C/O}, \texttt{Dynesty} \citep{Speagle2020}, emcee \citep{emcee2013}, \eureka \citep{Eureka2022}, ExoCTK \citep{exoctk}, \texttt{Forecaster} \citep{Chen2017}, IPython \citep{ipython}, \texttt{jwst} \citep{jwstpipeline2022}, Matplotlib \citep{matplotlib}, NumPy \citep{numpy, numpynew}, \texttt{PICASO} \citep{Batalha2019}, \texttt{POSEIDON} \citep{MacDonald2017}, PyMC3\citep{Salvatier2016}, SciPy \citep{scipy}, \texttt{smarter} \citep{Lustig-Yaeger2022}.

\bibliography{bibliography}

\end{document}